# Enhanced weakly-compressible MPS method for violent free-surface flows: Role of particle regularization techniques

Mojtaba Jandaghian[a], Abdelkader Krimi[a], Amir Reza Zarrati[b], Ahmad Shakibaeinia[a,c*]

[a] *Department of Civil, Geological and Mining Engineering, Polytechnique Montréal, Montréal, Canada*
[b] *Department of Civil and Environmental Engineering, Amirkabir University of Technology, Tehran, Iran*
[c] *Canada Research Chair in Computational Hydrosystems*

*\*ahmad.shakibaeinia@polymtl.ca*

**Abstract**

This paper develops a consistent particle method for capturing the highly non-linear behavior of violent free-surface flows, based on an Enhanced Weakly Compressible Moving Particle Semi-implicit (EWC-MPS) method. It pays special attention to the evaluation and improvement of two particle regularization techniques, namely, pairwise particle collision (PC) and particle shifting (PS). To improve the effectiveness of PC in removing noisy pressure field, and volume conservation issue of PS, we propose and evaluate several enhancements to these techniques, including a novel dynamic PC technique, and a consistent PS algorithm with new boundary treatments and additional terms (in the continuity and momentum equations). Besides, we introduce modified higher-order and anti-symmetric operators for the diffusive and shear force terms. Evaluation of the proposed developments for violent free-surface flow benchmark cases (2D dam-break, 3D water sloshing, and 3D dam-break with an obstacle) confirms an accurate prediction of the flow evolution and rigid body impact, as well as long-term stability of the simulations. The dynamic PC reduces pressure noises with low energy dissipation, and the consistent PS conserves the volume even for extreme deformations. Comparing the role of these new particle regularization techniques demonstrates the effectiveness of both in assuring the uniformity of the particle distribution and pressure fields; nevertheless, the implementation of PS is found to be more complex and time-consuming, mainly due to its need for free surface detection and boundary treatment with many tuning parameters.

**Graphical Abstract**

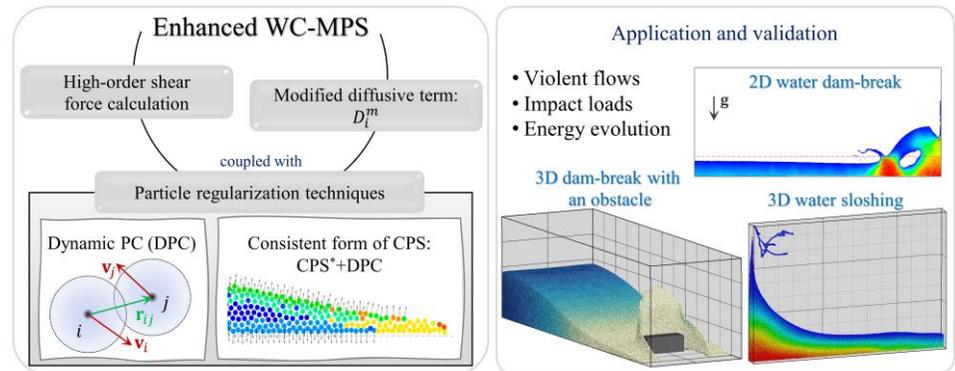

## 1. Introduction

Continuum-based particle methods have been developed and applied for solving many scientific and engineering problems (see e.g., [1-4]). The Lagrangian nature of these approaches has attracted many research projects involving complex fluid flow physics. In contrast to the traditional Eulerian methods identified with semi- or fixed grids, the moving particles carrying the flow and material properties provide an effective computational tool for capturing extreme flows and interface deformations. Nevertheless, the numerical stability and convergence of these particle methods are still open problems that require robust stabilization techniques (e.g., [5, 6]).

Smoothed Particle Hydrodynamics (SPH) and Moving Particle Semi-implicit (MPS) (initially developed by Gingold and Monaghan [7] and Koshizuka and Oka [8], respectively) are two popular particle methods for fluid flow problems. Shakibaeinia and Jin [9] introduced the Weakly Compressible MPS (WC-MPS) method, with the equation of state being the function of the particle number density. This method has been the focus of many studies, particularly for simulating hydro-environmental problems (e.g., see [10, 11] for free-surface and multiphase fluid flows, and [12, 13] for sediment dynamic modeling). Jandaghian and Shakibaeinia [14] recently proposed a new version of the WC-MPS approach, denoted as the Enhanced WC-MPS (EWC-MPS). The EWC-MPS method, as a continuity-based model, employs a diffusive term and particle regularization techniques to reduce/eliminate numerical instability. In the present study, we further improve this method and investigate its stability and convergence for the long-term modeling of violent free-surface flows by introducing several new enhancements.

Numerical instabilities in particle methods mainly arise from errors due to the discretization methods and the Lagrangian motion, which lead to non-uniform particle distributions recognized as noisy pressure fields, particle clustering, and tensile instability. These drawbacks of the particle methods not only cause numerical instabilities but also affect the modeling accuracy, particularly in regions with high-pressure and/or tension forces and extreme flow deformations [15]. To decrease the high-frequency unphysical pressure fluctuations, particle methods implement higher-order operators (e.g., [16-18]) and artificial viscosity/diffusion terms (e.g., [19, 20]). Moreover, particle regularization techniques have been developed to assure uniform particle distributions and overcome the common pairing instabilities. Among these regularization techniques, the pair-wise Particle Collision (PC) (e.g., [11, 21, 22]), the Particle Shifting (PS) (e.g., [5, 23-26]), and the transport-velocity formulations (e.g., [27-29]) approaches have shown to be promising in surmounting these instabilities by regulating the particles.

Unphysical high-frequency pressure noises, as the common numerical issue of the weakly compressible SPH and MPS methods, originate from the particle approximation and the kernel truncation errors [14, 20]. To reduce these pressure fluctuations, Monaghan and Gingold [19] inserted an artificial viscosity term into the momentum equation of the weakly compressible SPH method for shock simulations. The artificial viscosity term has shown to be an effective numerical technique for eliminating the pressure fluctuations and ensuring numerical stability, and is being implemented in well-developed SPH models (e.g., see [5, 30-32]). Molteni and Colagrossi [20] introduced a novel diffusive term defined as the Laplacian of the density field in the continuity equation to surmount the pressure oscillations in the weakly compressible SPH method. Antuono et al. [33] proposed the convergent form of this diffusive term using high-order Laplacian and gradient operators. In the context of the weakly compressible MPS, Jandaghian and Shakibaeinia [14] derived a new version of this diffusive term as a function of the particle number density for free-surface flows. Furthermore, high-order approximation operators have been employed to increase the accuracy of the numerical calculations and remove the pressure noises [17]. Oger et al. [16] improved the SPH method through a renormalization-based formulation for increasing the accuracy of the gradient approximations. Similarly, Khayyer and Gotoh [17] represented a first-order accuracy pressure gradient equation in the incompressible MPS framework. Duan et al. [18] implemented a second-order corrective matrix and showed that in their incompressible MPS model the stabilization errors are more dominant compared to the truncation errors. Jandaghian and Shakibaeinia [14] by comparing a conservative pressure gradient operator versus the non-conservative high-order pressure force illustrated that respecting the conservation properties of the system plays an important role in the numerical stability of the weakly compressible MPS method.

To increase the numerical stability, the particle methods adopt the PC technique developed based on the concept of the momentum transfer in the collision of physical solid or gas particles. This approach utilizes empirical coefficients to indirectly apply a repulsive force between a pair of colliding particles. Lee et al. [21] represented the PC technique as a particle regularization technique in the MPS formulations for simulating violent free-surface flows and impact loads. Shakibaeinia and Jin [12] developed the PC formulation for simulating sediment dynamic problems in the context of the multiphase weakly compressible MPS method. Further, Shakibaeinia and Jin [11] extended the application of the PC method to high-density ratio multiphase flows, and Xu and Jin [22] validated this technique as the particle stabilization approach of the weakly compressible MPS method for free-surface flows involving impact events.

Moreover, PS techniques aim at eliminating particle-clustering and tensile instability by moving particles to the area with less particle concentration. In the context of the incompressible SPH method, Xu et al. [23] originally implemented the PS method for solving internal flows. Lind et al. [24] represented the PS approach based on Fick's law of diffusion and corrected the PS vector for free-surface flows. Shadloo et al. [34] adopted the Lennard-Jones repulsive force to derive the PS vector and validated the improved SPH method for simulating rapid fluid flows over solid bodies. Zainali et al. [35] employed this PS formulation for multiphase incompressible SPH simulations. Khayyer et al. [25] proposed an optimized version of the PS scheme to enhance the performance of the incompressible SPH method for interfacial flows. Mokos et al. [36] developed the PS technique for multiphase violent fluid flows in the weakly compressible SPH framework. Duan et al. [26] implemented the PS technique within an incompressible multiphase MPS model considering the free-surface corrections of the PS vector. Sun et al. [37] appended the PS technique to the weakly compressible SPH method with special treatments of particles in the free-surface region. Further, Sun et al. [5] developed the consistent form of this PS method by including additional diffusive terms of the PS transport velocity within the governing equations. Jandaghian and Shakibaeinia [14] proposed the corrected-PS algorithm coupled with the PC technique (i.e., the hybrid method denoted as CPS+PC) for their enhanced weakly compressible MPS method, and the developed model was shown to be effective and essential for avoiding unphysical fluid fragmentations.

Violent free-surface flows are particularly challenging for the well-developed particle methods for capturing long-term mechanical behaviors (see e.g., [29, 38]). These complex flows are usually associated with breaking waves, impact events, and water sloshing, and are characterized by high

Reynolds numbers, highly non-linear deformations, and fluid-fluid and fluid-solid impacts [30]. Hence, increasing stability and reducing the approximation errors of the particle methods while conserving the global momentum/energy of the system become critical issues for simulating these problems. For this purpose, in addition to employing higher-order formulations and diffusive terms, eliminating the particle-pairing instabilities requires rigorous particle regularization techniques [5]. The PS techniques (e.g., [36, 38]), transport-velocity algorithms (e.g., [29, 39]), and the PC method (e.g., [21, 22]) have been specifically developed and utilized to surmount the numerical instability of violent free-surface flows. Nevertheless, despite their success in improving the particle distributions, many existing particle-shifting methods and the associated free-surface detection algorithms introduce some new challenges in simulating such complex flows (e.g., [6, 14]). Particularly in weakly compressible particle methods, where the potential energy is dominant and several breaking events occur (e.g., water dam break and sloshing problems), the continuous shifting of the particles leads to an unphysical expansion of the fluid field [5, 32]. Further developments to overcome these numerical issues involve implementing more accurate particle classification algorithms and treatment of the free-surface particles (e.g., [6, 25, 32]). Moreover, including some additional advection terms in the governing equations may dissipate the excessive potential energy inserted via the particle-shifting technique [5]. On the other hand, the conservative PC technique has advantages over PS because of its simplicity and being free from many exceptions and boundary treatments; however, by using empirical coefficients, this technique has been shown to be less effective for highly violent flows (see e.g., [21, 22]).

The main objective of this paper is to propose a robust numerical tool based on the EWC-MPS method for simulating the long-term stability of violent free-surface flows while predicting the system's global energy. To accomplish that, we propose a conservative form of the three-dimensional EWC-MPS method by employing a higher-order gradient and Laplacian operators (Section 3.1). Moreover, we implement a modified diffusive term in the MPS framework (based on the formulation of [33]) to reduce the high-frequency pressure fluctuations and kernel truncation errors. Regarding the particle regularization techniques, we investigate the role and impact of PC and PS techniques on stability, accuracy, and conservation properties, then to address the issues with these techniques we propose and evaluate a new dynamic PC approach (Section 3.2.1) and a consistent form of the CPS technique (Section 3.2.2). By simulating and validating two-dimensional dam-breaks and three-dimensional water sloshing and obstacle impacts, we investigate and compare the performance of the proposed developments for the long-term stability of relevant problems (Section 4).

## 2. Governing Equations

The mathematical model governs the fluid flows through the momentum and mass conservation laws. The Lagrangian form of the governing equations in the continuum mechanics is as follows [40]:

$$\begin{cases} \dfrac{D\rho}{Dt} = -\rho \nabla \cdot \mathbf{v} \\ \rho \dfrac{D\mathbf{v}}{Dt} = -\nabla p + \rho \mathbf{F} + \nabla \cdot \boldsymbol{\tau} \\ \dfrac{D\mathbf{r}}{Dt} = \mathbf{v} \end{cases} \quad (1)$$

In this system, the continuity equation updates the fluid density, $\rho$, by the divergence of the velocity, $\nabla \cdot \mathbf{v}$, stated as the mass volume expansion rate. The momentum equation updates the velocity vector, $\mathbf{v}$, used to move the position of the material points, $\mathbf{r}$. The gradient of pressure, $\nabla p$, the divergence of the shear stress tensor, $\nabla \cdot \boldsymbol{\tau}$, and the body force, $\rho \mathbf{F}$, represent the interaction forces per unit volume. With the weak compressibility assumption, the pressure, $p$, is a function of the density as a barotropic fluid given by the Equation of State (EOS) (i.e., $p = f(\rho)$). The shear force, $\nabla \cdot \boldsymbol{\tau}$, introduces the laminar and turbulence properties of the viscous fluid within the numerical model. For a Newtonian incompressible fluid (with $\nabla \cdot \mathbf{v} \simeq 0$), the shear stress tensor is given as a function of the dynamic viscosity, $\mu$, and the strain rate tensor of the fluid phase (i.e., $\boldsymbol{\tau} = 2\mu \mathbf{E}$ where $\mathbf{E} = 0.5[(\nabla \mathbf{v}) + (\nabla \mathbf{v})^T]$).

## 3. The EWC-MPS model

### 3.1 The discrete system with the conservation properties

The particle methods discretize the computational domain, $\Omega$, into particles that represent the fluid phase, $\Omega_f$, and the solid walls, $\Omega_s$ (i.e. $\Omega = \Omega_f \cup \Omega_s$). In the EWC-MPS method using the symbolic notation, the approximation operator, $\langle \ \rangle$, forms the differential equations (1) for a target particle $i$ into:

$$\begin{cases} \dfrac{1}{n_i}\dfrac{Dn_i}{Dt} = -\langle \nabla.\mathbf{v}\rangle_i + D_i^m \\ \dfrac{D\mathbf{v}_i}{Dt} = -\dfrac{1}{\rho_i}\langle \nabla p\rangle_i + \mathbf{F}_i + \dfrac{1}{\rho_i}\langle \nabla.\boldsymbol{\tau}\rangle_i \\ \dfrac{D\mathbf{r}_i}{Dt} = \mathbf{v}_i \end{cases} \quad (2)$$

in which the model calculates the particle number density, $n_i$, (as the substitute for the density, $\rho_i$) via the continuity equation supplied with the modified diffusive term, $D_i^m$ [14]. The momentum equation estimates the interparticle forces (i.e., the pressure force, $\langle \nabla p\rangle_i$, the shear force, $\langle \nabla.\boldsymbol{\tau}\rangle_i$, and the body force, $\rho_i \mathbf{F}_i$ per unit volume) for updating the velocity, $\mathbf{v}_i$, and eventually the position of the particle, $\mathbf{r}_i$, while the model keeps the fluid density constant in the momentum equation (i.e., $\rho_i = \rho_0$: the reference density of the fluid phase; for water, $\rho_0 = 1000\ kg/m^3$), the equation of state updates the pressure field as a function of $n_i$ with:

$$p_i = \dfrac{c_0^2 \rho_0}{\gamma}\left(\left(\dfrac{n_i}{n_0}\right)^\gamma - 1\right) \quad (3)$$

where $n_0$ is the normalization factor and $\gamma = 7$ [9]. With the artificial speed of sound, $c_0$, being considerably larger than the maximum velocity corresponding to the problem, $\|\mathbf{v}\|_{max}$, (e.g., $c_0 \geq 10\|\mathbf{v}\|_{max}$), the equation of state limits the compressibility of the fluid phase to 1%, as the Mach number ($Ma$) remains less than 0.1. The kernel function, $W(r_{ij}, r_e)$, defines the interaction of particle $i$ with its neighboring particles ($j \in N$) through the approximation procedure. The neighbor search algorithm updates the neighbor lists in each time step with the kernel influence radius, $r_e = 3.1l_0$, where $r_{ij} = \|\mathbf{r}_j - \mathbf{r}_i\| \leq r_e$ and $l_0$ is the initial particle distance. The conjugate form of the gradient and divergence operators for the pressure force and the velocity field, respectively, gives:

$$\begin{cases} \langle \nabla p\rangle_i = \dfrac{d}{n_0}\sum_{i \neq j}^{N}\left(n_i\dfrac{p_j}{n_j} + n_j\dfrac{p_i}{n_i}\right)\dfrac{\mathbf{e}_{ij}}{r_{ij}}W_{ij} \\ \langle \nabla.\mathbf{v}\rangle_i = \dfrac{d}{n_0}\sum_{i \neq j}^{N}\left(\dfrac{n_j}{n_i}\right)\dfrac{\mathbf{v}_{ij}}{r_{ij}}.\mathbf{e}_{ij}W_{ij} \end{cases} \quad (4)$$

in which $\mathbf{e}_{ij}$ is the unit direction vector between particle $i$ and $j$ (i.e. $\mathbf{e}_{ij} = (\mathbf{r}_{ij} = \mathbf{r}_j - \mathbf{r}_i)/r_{ij}$), $W_{ij} = W(r_{ij}, r_e)$, and $d\ (= 1, 2, 3)$ is the number of space dimensions [14]. The normalization factor, $n_0$, is calculated at the initial stage ($t = 0$) by the kernel summation over the fluid particles with the complete kernel support (i.e., $n_0 = \max(\sum_{i \neq j}^N W_{ij})$ at $t = 0$). We employ the rational kernel function [9] for all the approximated terms of the governing equation (2).

Employing the conjugate form of the approximation operators in (4) ensures that the particle system preserves the total energy of the numerical model (in the absence of the shear and body forces and the diffusion terms) [14, 41]. Further, this model conserves the linear and angular momentums, since the pressure force is antisymmetric (with respect to the particle indexes) and with the same direction as the $\mathbf{r}_{ij}$ vector. In the presence of viscous stresses, we should approximate the shear force, $\langle \nabla . \boldsymbol{\tau} \rangle_i$, with an antisymmetric formulation in order to still conserve the linear momentum of the system. To do so, in this study we implement the following divergence operator:

$$\langle \nabla . \boldsymbol{\tau} \rangle_i = \frac{d}{n_0} \sum_{i \neq j}^N \left( \frac{n_i}{n_j} \boldsymbol{\tau}_j + \frac{n_j}{n_i} \boldsymbol{\tau}_i \right) . \frac{\mathbf{e}_{ij}}{r_{ij}} W_{ij} \qquad (5)$$

derived based on the formulation by Jandaghian and Shakibaeinia [14] proposed for the conservative pressure force in Eq. (4). In this equation, the shear stress tensor of particle $i$ is given by $\boldsymbol{\tau}_i = 2\mu_i \mathbf{E}_i$, where $\mathbf{E}_i$ is the deviatoric part of the strain rate tensor estimated by $0.5(\langle \nabla \mathbf{v} \rangle_i + \langle \nabla \mathbf{v} \rangle_i^T)$ for incompressible flows. Furthermore, we implement a turbulence model to simulate unresolved fluid motions smaller than the filter size [42]. Based on the Sub Particle Scale (SPS) turbulence model, we add the turbulent viscosity, $\mu_{t_i}$, to the dynamic viscosity of the fluid phase (i.e., $\mu_i = \mu + \mu_{t_i}$ and for water $\mu = 10^{-3}\ Pa.s$), where $\mu_{t_i} = \rho_0 (C_s r_e)^2 \sqrt{4II_{E_i}}$ and $II_{E_i} = 0.5 \mathbf{E}_i : \mathbf{E}_i$ is denoted as the second main invariant of the strain rate tensor and $C_s$ is the Smagorinsky coefficient set to 0.12 in this study [13, 43].

Even with Eq. (5) being an antisymmetric equation with respect to the particle indexes, the direct methods are incapable of confirming the conservation of the angular momentum within the discrete system [44]. Considering the variational procedure, Bonet and Lok [44] showed that the numerical model preserves the angular momentum if the potential energy remains invariant as the rigid body rotates. This is achieved by implementing a gradient operator that correctly estimates the true velocity gradient, $\mathbf{W}$, of an angular velocity vector, $\boldsymbol{\omega}$, where $\mathbf{v}_i = \boldsymbol{\omega} \times \mathbf{r}_i$ (i.e., ideally $\langle \nabla \mathbf{v} \rangle_i = \mathbf{W}$) [44]. In the MPS framework, the gradient of this velocity field is approximated via:

$$\langle \nabla \mathbf{v} \rangle_i = \frac{d}{n_0} \sum_{i \neq j}^{N} \frac{\mathbf{W}\mathbf{r}_{ij}}{r_{ij}} \otimes \mathbf{e}_{ij} W_{ij} \tag{6}$$

in which $\otimes$ is the outer product of two vectors. Therefore, the condition that assures the preservation of the related potential energy will be:

$$\frac{d}{n_0} \sum_{i \neq j}^{N} \frac{\mathbf{r}_{ij}}{r_{ij}} \otimes \mathbf{e}_{ij} W_{ij} = \mathbf{I} \tag{7}$$

which is required for estimating the true velocity gradient ($\mathbf{I}$ being the identity matrix). Satisfying this condition leads to a corrective matrix identical to $\mathbf{C}_i$, given as:

$$\mathbf{C}_i = \left( \frac{d}{n_0} \sum_{i \neq j}^{N} \frac{\mathbf{r}_{ij}}{r_{ij}} \otimes \mathbf{e}_{ij} W_{ij} \right)^{-1} \tag{8}$$

which ensures the exact calculation of the gradient of any linear velocity field [16, 17, 44]. Accordingly, we include the corrective matrix into the gradient operator of the velocity vector as follows:

$$\langle \nabla \mathbf{v} \rangle_i^C = \frac{d}{n_0} \sum_{i \neq j}^{N} \frac{\mathbf{v}_{ij}}{r_{ij}} \otimes (\mathbf{C}_i \mathbf{e}_{ij}) W_{ij} \tag{9}$$

which is used in the shear force calculations (i.e., in the $\boldsymbol{\tau}_i$ and $\mu_{t_i}$ terms) and $\mathbf{v}_{ij} = \mathbf{v}_j - \mathbf{v}_i$. By implementing this first-order accuracy operator, we aim to conserve the angular momentum of the system (similar to the incompressible SPH method proposed by Khayyer et al. [42]).

Based on the flow equations in the MPS framework, Jandaghian and Shakibaeinia [14] introduced the diffusive term into the continuity-based WC-MPS model to eliminate high-frequency pressure fluctuations. We modify the standard diffusive term by employing a higher-order Laplacian operator to implement its convergent form (like the formulation proposed by Antuono et al. [33] for the $\delta$-SPH models). This eliminates kernel truncation errors at the free surface and in the vicinity of the solid boundary, converging the diffusive term over the fluid domain, which is required to obtain a stable pressure field for long-term simulations [33]. In the context of the WC-MPS method, the modified the diffusive term forms to:

$$D_i^m = \left(\delta_{MPS}\frac{\Delta t c_0^2}{n_0}\right)\frac{2d}{n_0}\sum_{i\neq j}^{N}\left[(n_j - n_i) - \frac{1}{2}[\langle\nabla n\rangle_i^C + \langle\nabla n\rangle_j^C].\mathbf{r}_{ij}\right]\frac{W_{ij}}{r_{ij}^2} \tag{10}$$

in which the time step of the calculation, $\Delta t$, the reference sound speed, $c_0$, the normalization factor, $n_0$, and the adjusting coefficient, $0 < \delta_{MPS} \leq 1$ determine the strength of the numerical diffusion (in this study, $\delta_{MPS} = 0.7$ for all test cases). Using the higher-order gradient equation produces the gradient of the particle number density, as follows:

$$\langle\nabla n\rangle_i^C = \frac{d}{n_0}\sum_{i\neq j}^{N}\frac{n_j - n_i}{r_{ij}}(\mathbf{C}_i\mathbf{e}_{ij})W_{ij} \tag{11}$$

where the corrective matrix, $\mathbf{C}_i$, imposes the first-order accuracy on the gradient operator. It should be noted that the modified diffusive terms conserve the total mass of the system, as equations (8), (10), and (11) are limited to the fluid particles (i.e. $i, j \in \Omega_f$) and $\sum_i n_i D_i^m V_i = 0$, $V_i$ is the volume of the particle.

## 3.2 Particle regularization techniques for violent free-surface flows

### 3.2.1 Dynamic pair-wise Particle Collision technique: DPC

The idea behind the pair-wise Particle Collision (PC) approach originates from the momentum transfer between colliding physical solid or gas particles in the normal direction [11, 45]. In the particle methods (e.g., [11, 21, 22]), the standard PC technique modifies the velocity of fluid particles penetrating each other considering the collision velocity, a constant coefficient of restitution, and a threshold as the collision distance ($\mathbf{v}_{ij}^{Coll}$, $\varepsilon$ and $\theta l_0$, respectively) through $\mathbf{v}_i' = \mathbf{v}_i + \delta\mathbf{v}_i$ as:

$$\delta\mathbf{v}_i = \sum_{i\neq j}^{N}\left(\frac{1+\varepsilon}{2}\right)\mathbf{v}_{ij}^{Coll}$$

in which: $\tag{12}$

$$\mathbf{v}_{ij}^{Coll} = \begin{cases}(\mathbf{v}_{ij}.\mathbf{e}_{ij})\mathbf{e}_{ij} & \text{for } r_{ij} \leq \theta l_0 \text{ and } \mathbf{v}_{ij}.\mathbf{r}_{ij} < 0 \\ 0 & \text{Otherwise}\end{cases}$$

and $\varepsilon$ specifies the kinetic energy variation of the particles after the collision and $\theta$ is a non-dimensional threshold coefficient usually set to $0.8 - 0.9$ [11, 45]. The time integration algorithm dispositions the particle with the post-collision velocity $\mathbf{v}_i'$ [11]. This velocity modification ignores the inter-particle penetration of other neighboring particles, $j$, that are not approaching particle $i$ (i.e., for $\mathbf{v}_{ij}.\mathbf{r}_{ij} \geq 0 \rightarrow$ $\mathbf{v}_{ij}^{Coll} = 0$ even if $r_{ij} < l_0$). Moreover, the constant coefficient and the predefined threshold of collision distance restrict the effectiveness of the PC method for dealing with extreme flow deformations.

Here, we propose a new formulation of the PC technique to resolve the numerical instabilities associated with particle-clustering in violent free-surface flows. The developed algorithm employs variable coefficients and the local gradient of a dynamic background pressure to update the transport velocity. This particle regularization technique, hereinafter denoted as the Dynamic pair-wise Particle Collision (DPC) technique, modifies the velocity by:

$$\delta \mathbf{v}_i = \left( \sum_{i \neq j}^{N} \kappa_{ij} \mathbf{v}_{ij}^{Coll} - \frac{\Delta t}{\rho_0} \sum_{i \neq j}^{N} \phi_{ij} \frac{p_{ij}^b}{r_{ij}} \mathbf{e}_{ij} \right) \tag{13}$$

where:

$$(\mathbf{v}_{ij}^{Coll}, \phi_{ij}) = \begin{cases} \left((\mathbf{v}_{ij} \cdot \mathbf{e}_{ij})\mathbf{e}_{ij}, 0\right) & \text{If } \mathbf{v}_{ij} \cdot \mathbf{r}_{ij} < 0, \\ (0, 1) & \text{otherwise} \end{cases}$$

acting between the fluid particles (i.e., $i, j \in \Omega_f$). For pairs of interpenetrated particles still approaching each other, i.e., $\mathbf{v}_{ij} \cdot \mathbf{r}_{ij} < 0$, only the collision term (i.e., the first term on the right-hand side) modifies their velocity vectors with a variable coefficient, given as $0 \leq \kappa_{ij} \leq 1$, and a binary multiplier set to $\phi_{ij} = 0$. Otherwise, when they are receding from each other, the second term, as the local gradient of dynamic background pressure, $p_{ij}^b$, activates an artificial repulsive force between the two target particles by setting $\phi_{ij} = 1$.

This new formulation (i.e., Eq. (13)) covers the different conditions that demonstrate inter-particle penetration, i.e., particle-clustering (shown in Figure 1). By calculating the velocity variation of all the fluid particles, the proposed DPC technique updates the velocities and positions of the fluid particles as follows:

$$\begin{cases} \mathbf{v}_i' = \mathbf{v}_i + \delta \mathbf{v}_i \\ \mathbf{r}_i' = \mathbf{r}_i + \Delta t \delta \mathbf{v}_i \end{cases} \tag{14}$$

We define the dynamic background pressure by the local pressure of the particles and the inter-particle distance as follows:

$$\begin{cases} p_{ij}^b = \tilde{p}_{ij}\chi_{ij} \\ \tilde{p}_{ij} = \max\left(\min(\lambda|p_i + p_j|, \lambda p_{max}), p_{min}\right) \\ \chi_{ij} = \sqrt{\dfrac{W(r_{ij}, l_0)}{W(0.5l_0, l_0)}} \end{cases} \qquad (15)$$

where $\lambda$ is a non-dimensional coefficient (typically with a constant value of 0.1 ~ 0.2). This formulation limits the background pressure by the expected minimum and maximum values of the pressures (i.e., $p_{min}$ and $\lambda p_{max}$, respectively, considered as positive constants in the problem). The minimum pressure keeps the repulsive force non-zero, while the maximum pressure eliminates large velocity variations (particularly in regions where flow impacts solid walls or forms a plunging wave). The proposed dynamic background pressure (within the repulsive term) increases the stability particularly close to the free-surface where attractive forces may become dominant through the conservative pressure force. In this study, we set $\lambda = 0.2$ for all benchmark cases and assign $p_{min}$ and $p_{max}$ considering the minimum hydrostatic pressure on the free-surface and the maximum impact load on the walls, respectively. In test cases where the gravitational force is not dominant, these parameters can be determined according to the analytical solution and/or the local particle pressure. The kernel function, with a smoothing length equal to the initial particle distance, adjusts the strength of the repulsive force through $\chi_{ij}$, defined as a non-dimensional variable. This function increases as the particle distance decreases and becomes greater than the unit where $r_{ij} < 0.5l_0$. Similarly, we utilize $\chi_{ij}$ as the variable coefficient of the collision term considering the following conditions:

$$\kappa_{ij} = \begin{cases} \chi_{ij} & 0.5l_0 \leq r_{ij} < l_0 \\ 1 & r_{ij} < 0.5l_0 \end{cases} \qquad (16)$$

by which the magnitude of the velocity variation decreases smoothly with the particle separation. Further, the latter condition avoids excessive transfer of linear momentum between the colliding particles (i.e., with $\kappa_{ij} = 1.0$, the coefficient of restitution, $\varepsilon$, reaches its maximum value which is unit). Through the dynamic coefficients, both the collision and repulsive terms only modify the velocities and positions considering the nearest-neighbor particles that penetrate the target particle (i.e., if $r_{ij} \geq l_0$, $W(r_{ij}, l_0) = 0 \rightarrow \chi_{ij} = 0$). Note that since Eq. (13) is an antisymmetric term, the proposed DPC still conserves the total linear momentum of the system (i.e., $\sum_i m\delta\mathbf{v}_i = 0$ with the constant mass of particles $m$).

The repulsive term in Eq. (13) regulates particles based on the pressure gradient term of the momentum equation (i.e., $d\mathbf{v}/dt = -\nabla p/\rho_0$, similar to the generalized transport-velocity approach of Zhang et al. [28] and the artificial repulsive forces of Monaghan et al. [46] and Tsuruta et al. [47]). However, the proposed repulsive force coupled with the new collision term still acts as a pair-wise technique without the smoothing procedure over all the neighbor particles. Further, this term is antisymmetric, and the velocity is directly updated by Eq. (14) without implementing any additional term in the momentum equation (i.e., without $(\nabla.(\mathbf{v} \otimes \delta\mathbf{v}) - \mathbf{v}\nabla.(\delta\mathbf{v}))$ terms). It should be noted that the effectiveness of the DPC technique depends on the type of kernel function used for the $\chi_{ij}$ variable (e.g., the kernel functions with the singularity issues (like the rational kernel function [9]) may lead to large values of the repulsive force when $r_{ij} \to 0$). In this study, we use the Wendland kernel function [48], where $0 \leq \chi_{ij} \leq 2.31$. Figure 1 illustrates the numerical implementation of the DPC algorithm and plots the Wendland kernel and $\chi_{ij}$ functions.

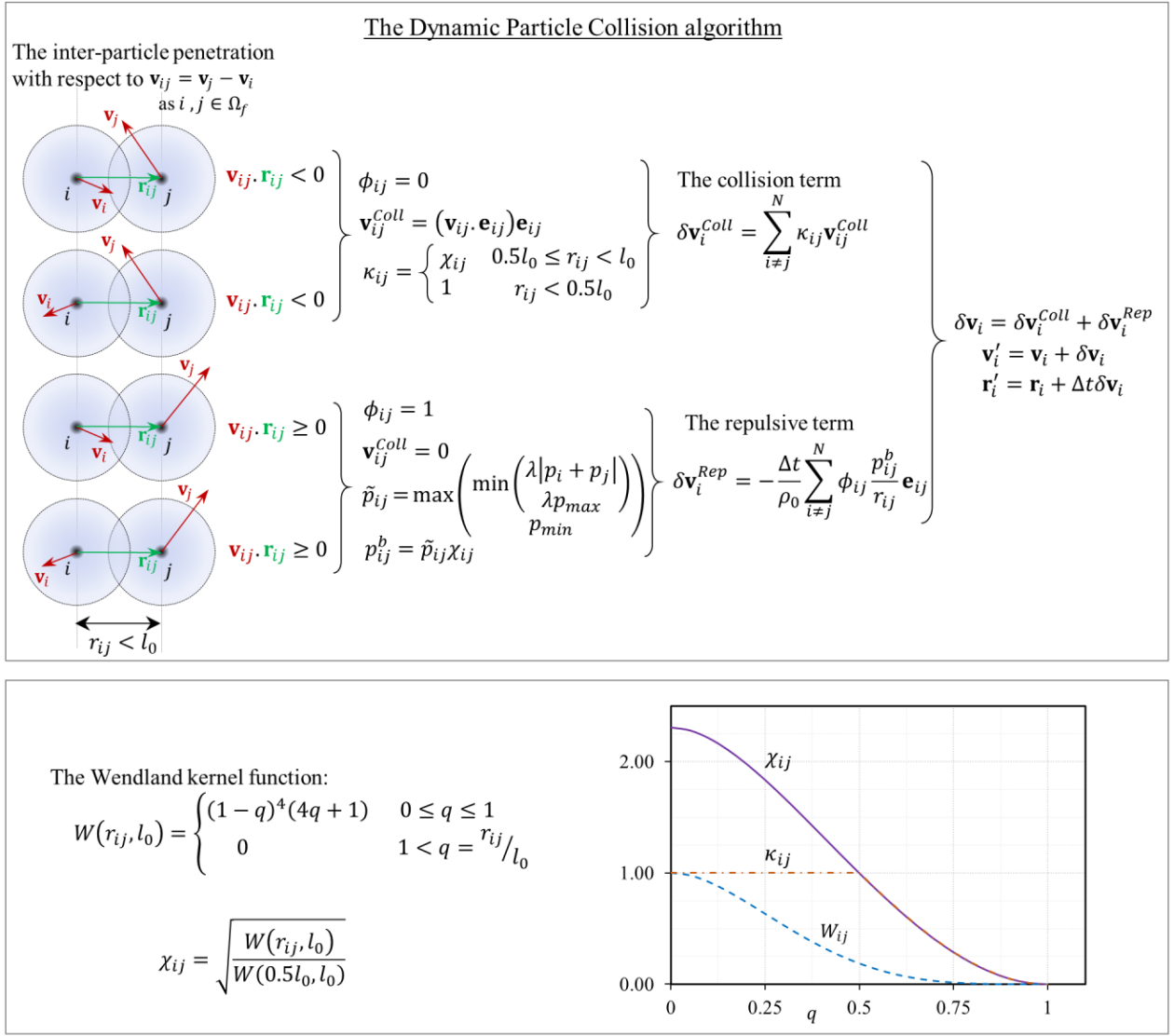

Figure 1- The Dynamic Particle Collision (DPC) algorithm consists of the proposed collision and repulsive terms and the dynamic coefficients defined by the Wendland kernel function.

### 3.2.2 A consistent form of the Corrected Particle Shifting technique: CPS*+DPC

The particle shifting (PS) algorithm regulates the distribution of fluid particles, thus eliminating the tensile instability and increasing the numerical accuracy and stability. Here, we propose and develop a consistent form of the corrected-PS algorithm (i.e., the hybrid approach originally proposed in [14] denoted as CPS+PC) for the EWC-MPS method. To that end, we implement a more restricted criterion for boundary treatments of the particle-shifting vector. We also re-derive the MPS formulation of the additional diffusive terms included in the continuity and momentum equations resulting from the particle-

*"Enhanced weakly-compressible MPS method for violent free-surface flows: Role of particle regularization techniques"*
*Jandaghian et al. (2021), ACCEPTED MANUSCRIPT, https://doi.org/10.1016/j.jcp.2021.110202*shifting velocity (based on the work of Sun et al. [5]). The former introduces new criteria to classify the fluid particles required for dealing with corners and extreme curvatures of the interfaces. The latter aims to include the effects of the gradients' advection due to the particle shifting transport-velocity within the governing equations. Eventually, coupling the new CPS algorithm with the DPC technique, we propose a consistent form of the corrected particle-shifting technique, denoted as CPS$^*$+DPC, for the free-surface flows.

- *Numerical treatment of the particle-shifting vector*

Here, the conservative particle shifting vector of the fluid particles, $i \in \Omega_f$, is given by:

$$\delta \mathbf{r}_i = -F\nabla C_i = -F\frac{d}{n_0}\sum_{i \neq j|i,j\in\Omega_f}^{N}\frac{C_i + C_j}{r_{ij}}\mathbf{e}_{ij}W_{ij} \tag{17}$$

where $j$ only belongs to the fluid particles, $\Omega_f$, and $F = Al_0^2 C_{CFL} Ma$ is the Fickian diffusion coefficient with $A$ and $C_{CFL}$ being a dimensionless number ($A = 2$ in this study) and the Courant condition coefficient of the explicit model, respectively. Unlike in [14], Eq. (17) excludes the contribution of the solid boundary particles to the particle-shifting of the fluid particles. This vector dispositions the fluid particles to the area with less particle concentration, quantified by $C_i = \sum_{i\neq j}^{N} W_{ij}/n_0$ as $j \in \Omega$. Thus, the CPS algorithm should modify/cancel the particle-shifting vector with the incomplete kernel support (i.e., the particles in the vicinity of the free-surface and solid boundaries) to eliminate false particle-shifting.

For this purpose, the new dynamic particle classification algorithm, in addition to the internal particles ($i \in \mathbb{I}$), the free-surface vicinity particles ($i \in \mathbb{B}_F$), the free-surface particles ($i \in \mathbb{F}$), and the external/splashed particles ($i \in \mathbb{E}$), identifies the wall vicinity particles ($i \in \mathbb{B}_w$) and the fluid particles that signify the local volume expansion ($i \in \mathbb{X}$). Note that these classes represent the entire fluid domain (i.e., $\Omega_f = \mathbb{I} \cup \mathbb{F} \cup \mathbb{B}_F \cup \mathbb{B}_w \cup \mathbb{X} \cup \mathbb{E}$), whereas the wall and ghost solid boundary particles are denoted as $\mathbb{W}$ and $\mathbb{G}$, respectively (i.e., $\Omega_s = \mathbb{W} \cup \mathbb{G}$). Based on the particle shifting correction proposed by Lind et al. [24] and Khayyer et al. [25], the false diffusion of the fluid particles toward the free surface should be avoided. In this study, with the fluid particles being categorized, the new CPS formulation modifies the particle-shifting vector through:

15© 2021. This manuscript version is made available under the CC-BY-NC-ND 4.0
license https://creativecommons.org/licenses/by-nc-nd/4.0/

$$\delta \mathbf{r}_i = \begin{cases} -F \nabla C_i, & i \in \mathbb{I} \\ -F[\alpha_i \mathbf{I} - \beta_i(\hat{\mathbf{n}}_i \otimes \hat{\mathbf{n}}_i)] . \nabla C_i, & i \in \mathbb{B}_\text{F} \cup \mathbb{B}_\text{w} \\ -F[\alpha_i \mathbf{I} - \beta_i(\mathbf{n}_i \otimes \mathbf{n}_i)] . \nabla C_i, & i \in \mathbb{F} \\ 0 & i \in \mathbb{E} \cup \mathbb{X} \end{cases} \quad (18)$$

in which **I** is the identity matrix. This formulation removes the normal component of the particle-shifting vector of the free-surface particles ($i \in \mathbb{F}$) and the particles identified in the vicinity of the free-surface and the solid boundary particles ($i \in \mathbb{B}_\text{F} \cup \mathbb{B}_\text{w}$) via their unit local normal vector, $\mathbf{n}_i$, and the non-local normal vectors $\hat{\mathbf{n}}_i$, respectively [14]. The binary coefficients $\alpha_i$ and $\beta_i$ (which are either equal to 1 or 0) in the new equation impose some restricting conditions on the particle-shifting vector of the free-surface and boundary particles (through Eqs. (24)-(27)).

Firstly, the algorithm distinguishes the free-surface particles and the regions with unphysical volume expansion (due to manipulation of the potential energy by the particle-shifting procedure). To do so, the model evaluates $C_i$ with the predefined limits and then searches for the neighboring particles, $j \in \Omega$, within the umbrella-shaped region: R1 and R2, defined in [14] as:

$$\begin{cases} \text{R1: } |\mathbf{r}_{ij}| \geq \sqrt{2} l_0 \text{ and } |\mathbf{r}_{jS}| < l_0 \\ \text{R2: } |\mathbf{r}_{ij}| < \sqrt{2} l_0 \text{ and } |\mathbf{n}_i . \mathbf{r}_{jS}| + |\mathbf{t}_i . \mathbf{r}_{jS}| < l_0 \end{cases} \quad (19)$$

where $S$ is at the distance $l_0$ from particle $i \in \Omega_f$ in the normal direction and the unit local normal vector of the fluid particles, $\mathbf{n}_i$, is given as:

$$\mathbf{n}_i = \frac{\mathbf{N}_i}{\|\mathbf{N}_i\|} \text{ as } \mathbf{N}_i = \sum_{i \neq j \in \Omega}^{N} \mathbf{e}_{ji} W_{ij}, \quad (20)$$

noting that the unit tangential vector, $\mathbf{t}_i$, satisfies $\mathbf{t}_i . \mathbf{n}_i = 0$ condition (see R1 and R2 illustrated in Figure 2). Fluid particles with incomplete kernel support and without a neighboring particle inside their search regions are identified as free-surface particles (i.e., if $0.4 \leq C_i \leq 0.9$ and $j \notin \text{R1} \cup \text{R2} \rightarrow i \in \mathbb{F}$). A particle that still satisfies the $0.4 \leq C_i \leq 0.9$ condition while finding a neighboring particle within the R1 ∪ R2 region not detected as a free-surface or a wall particle, indicates an unphysical local expansion of volume in the internal region. To reduce the accumulation of errors in the potential energy of the system, we cancel the shifting of these particles (i.e., if $0.4 \leq C_i \leq 0.9$ and $j \in \text{R1} \cup \text{R2}$ & $j \notin \mathbb{F} \cup \mathbb{W} \rightarrow i \in \mathbb{X}$ & $\delta\mathbf{r}_i = 0$). Further, for detecting particles that are detached from the main body of the fluid domain, i.e., $i \in \mathbb{E}$, we consider the number of surrounding neighbors, $N_i$, (as $r_{ij} \leq r_e$ with $j \in \Omega$) and $C_i$ satisfying the $C_i < 0.4$ or $N_i < 0.4 N_0$ condition (where $N_0$ is the maximum number of neighbors as the isotropic particle distribution forms (e.g., with $r_e = 3.1 l_0$ in the two- and three-dimensional problems, $N_0$ is set to 34 and 140, respectively). In this case, its particle-shifting vector is neglected due to the extreme kernel truncations and the insufficient number of neighboring particles.

In the next step, the fluid particles that contain free-surface particles within their neighbor list (i.e., $r_{ij} \leq (r_e - l_0)$ as $j \in \mathbb{F}$) are identified as free-surface-vicinity particles ($i \in \mathbb{B}_F$) [14]. For these particles, the non-local normal vector, $\hat{\mathbf{n}}_i$, defined by [14] as:

$$\hat{\mathbf{n}}_i = \frac{\hat{\mathbf{N}}_i}{\|\hat{\mathbf{N}}_i\|} \text{ as } \hat{\mathbf{N}}_i = \sum_{j \in \mathbb{F}}^{N} \mathbf{n}_j W_{ij} \text{ for } i \in \mathbb{B}_F \tag{21}$$

cancels the shifting vector in the direction normal to the free surface. Furthermore, the new algorithm considers fluid particles that interact with the wall particles (i.e., $r_{ij} \leq (r_e - l_0)$ as $j \in \mathbb{W}$) in the vicinity of the wall ($i \in \mathbb{B}_w$). The non-local normal vectors of the wall-vicinity particles are specified by:

$$\hat{\mathbf{n}}_i = \frac{\hat{\mathbf{N}}_{w_i}}{\|\hat{\mathbf{N}}_{w_i}\|} \text{ as } \hat{\mathbf{N}}_{w_i} = \sum_{j \in \mathbb{W}}^{N} \mathbf{n}_{w_j} W_{ij} \text{ for } i \in \mathbb{B}_w, \tag{22}$$

where the unit normal and normal vectors of the wall particles (i.e., $\mathbf{n}_{w_i}$ and $\mathbf{N}_{w_i}$, respectively) are given as follows:

$$\mathbf{n}_{w_i} = \frac{\mathbf{N}_{w_i}}{\|\mathbf{N}_{w_i}\|} \text{ as } \mathbf{N}_{w_i} = \sum_{i \neq j \in \mathbb{W} \cup \mathbb{G}}^{N} \mathbf{e}_{ij} W_{ij} \text{ for } i \in \mathbb{W}. \tag{23}$$

With the unit normal vector given by (22), particle-shifting toward the solid boundary is ignored through Eq. (18). Finally, the fluid particles that are not in the vicinity of the free-surface and solid boundaries (i.e., $i \notin \mathbb{F} \cup \mathbb{B}_F \cup \mathbb{B}_w$) and the kernel support is almost complete (i.e., $C_i > 0.9$) are labeled as internal particles.

Considering the local curvature of the surfaces and the direction of the shifting vector against the normal vectors, the algorithm determines the binary coefficients $\alpha_i$ and $\beta_i$. In case the shifting vector of a fluid particle located in the vicinity of the free surface is toward the interior domain, the algorithm treats the particle as an internal one, i.e.:

$$\text{for } i \in \mathbb{B}_F \to (\alpha_i, \beta_i) = \begin{cases} (1,0) & \text{if } \delta\mathbf{r}_i.\widehat{\mathbf{N}}_i \leq 0 \\ (1,1) & \text{otherwise} \end{cases} \quad (24)$$

where $\delta\mathbf{r}_i$ is given by Eq. (17). To deal with fluid particles that are detected inside both the $\mathbb{F} \cup \mathbb{B}_F$ and $\mathbb{B}_w$ regions, special treatment of the shifting vector becomes necessary. This occurs at the intersection of the free surface region and the solid boundary. If this fluid particle's shifting vector (given by Eq. (17)) is toward either its free-surface or wall-normal vectors, the algorithm ignores its shifting, i.e.:

$$\text{for } i \in (\mathbb{F} \cup \mathbb{B}_F) \cap \mathbb{B}_w \to$$
$$(\alpha_i, \beta_i) = \begin{cases} (1,0) & \text{if } [\delta\mathbf{r}_i.(\mathbf{N}_i \text{ or } \widehat{\mathbf{N}}_i) \leq 0 \text{ and } \delta\mathbf{r}_i.\widehat{\mathbf{N}}_{w_i} \leq 0] \\ (0,0) & \text{otherwise} \end{cases}. \quad (25)$$

Moreover, for large surface curvatures (such as the corners of the fluid domain), correction of the shifting vector with the normal vector in a direction tangential to the surface still leads to inaccurate particle-shifting. Therefore, we introduce a criterion for distinguishing regions with large curvatures in order to ignore their particle-shifting. To calculate the scalar value of curvature, we estimate the divergence of the unit normal vectors by:

$$\kappa_i = \langle \nabla.\mathbf{n} \rangle_i = \frac{d}{n_0} \sum_{i \neq j}^{N} \frac{\mathbf{n}'_j - \mathbf{n}'_i}{r_{ij}}.\mathbf{e}_{ij} W_{ij} \quad (26)$$

$$\text{for } i \in (\mathbb{F} \cup \mathbb{B}_F) \cup \mathbb{B}_w \text{ while } i \notin (\mathbb{F} \cup \mathbb{B}_F) \cap \mathbb{B}_w$$

in which, if $i \in \mathbb{F} \to \mathbf{n}'_i = \mathbf{n}_i$ and if $i \in (\mathbb{B}_w \text{ or } \mathbb{B}_F) \to \mathbf{n}'_i = \widehat{\mathbf{n}}_i$. Further, $\mathbf{n}'_j = \mathbf{n}_{j \in \mathbb{F}}$ and $\mathbf{n}'_j = \mathbf{n}_{w_{j \in \mathbb{W}}}$ for $i \in (\mathbb{F} \cup \mathbb{B}_F)$ and $i \in \mathbb{B}_w$, respectively, while the neighbor particles that belong to intersecting free-surface and wall boundary particles are ignored (i.e., $j \in (\mathbb{W} \cup \mathbb{F}) \mid j \notin (\mathbb{F} \cap \mathbb{B}_w)$). With the scalar value of the curvature, $\kappa_i$, exceeding the predefined limit, $\kappa_{max}$, the binary coefficients set the shifting vector to null, i.e.:

$$i \in \mathbb{F} \cup (\mathbb{B}_F \text{ or } \mathbb{B}_w) \to (\alpha_i, \beta_i) = \begin{cases} (0,0) & \text{if } \kappa_i \geq \kappa_{max} \\ (1,1) & \text{otherwise} \end{cases} \quad (27)$$

wherein both two- and three-dimensional problems $\kappa_{max}$ is adjusted to 5.0, validated for the rational kernel function with $r_e = 3.1l_0$. This condition entirely cancels the shifting of free-surface and boundary particles in regions where the normal vectors are non-uniform. It should be highlighted that the only intersection between the particle categories occurs with fluid particles belonging to both the $\mathbb{F} \cup \mathbb{B}_F$ and $\mathbb{B}_w$ regions, specifically treated by Eq. (25); whereas no other identified regions overlap.

Altogether, the new particle-shifting vector, i.e., Eq. (18), modifies the positions of the fluid particles as $\mathbf{r}'_i = \mathbf{r}_i + \delta\mathbf{r}_i$. To avoid extreme deviation of the flow variables, we limit the magnitude of the particle-shifting vector by [24]:

$$\delta\mathbf{r}_i = \min(0.1\|\mathbf{v}_i\|\Delta t, 0.1 l_0, \|\delta\mathbf{r}_i\|) \frac{\delta\mathbf{r}_i}{\|\delta\mathbf{r}_i\|}. \qquad (28)$$

Figure 2 summarizes the dynamic particle classification algorithm with the special treatment of the free-surface and boundary particles by the new particle-shifting formulation. Also, this figure shows the particles' categories and the unit-normal vectors on the dam-break test case as an example (with $H/l_0 = 100$, $H$ being the initial height of the water column).

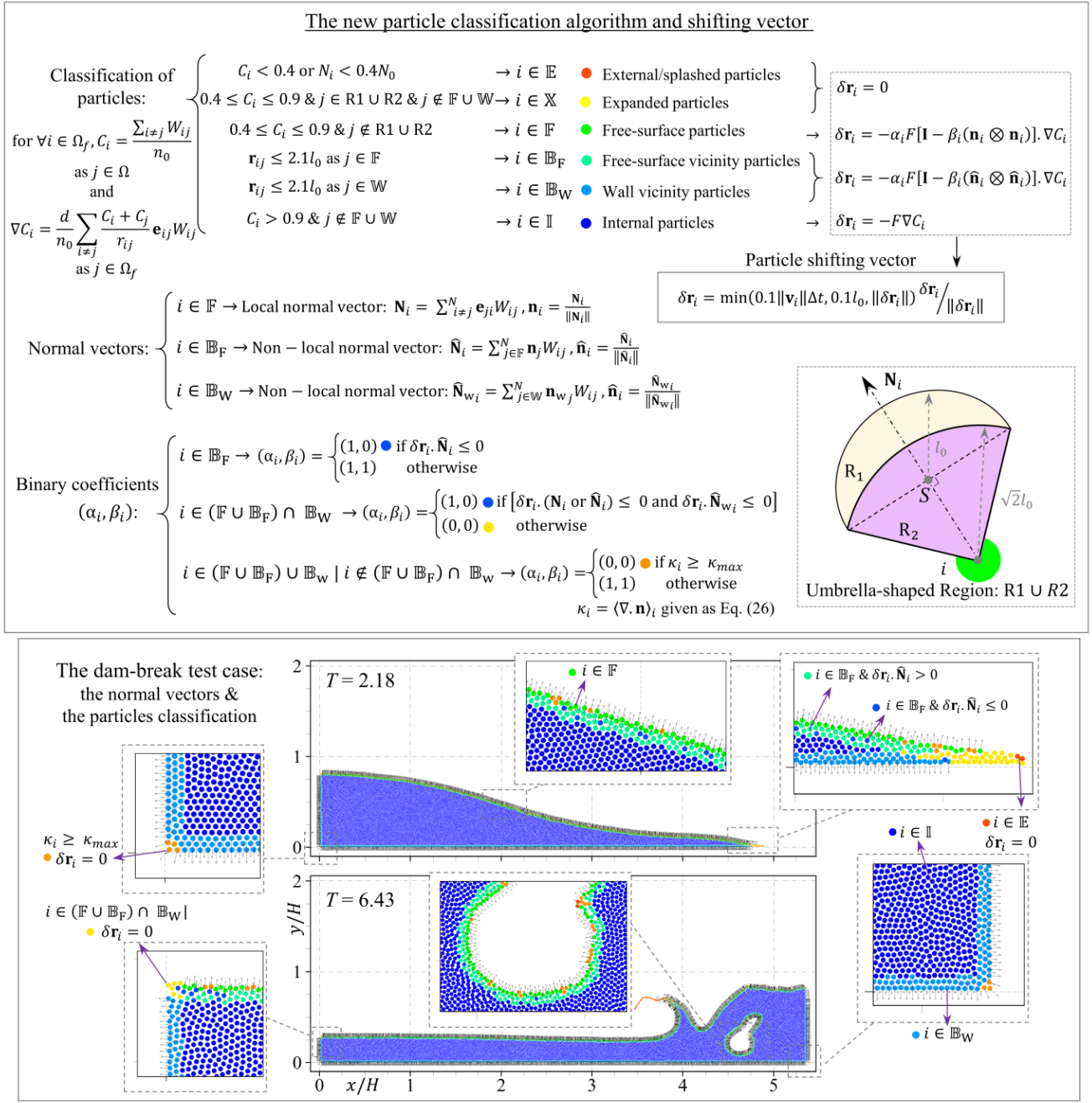

Figure 2- The new particle classification algorithm and special treatments of the boundary particles for the corrected particle-shifting vector. The umbrella-shaped region used for identifying the free-surface particle is illustrated (i.e., R1 and R2 given by Eq. (19)). The results of the two-dimensional (2D) dam-break test case (with $H/l_0 = 100$) show the estimated normal vectors and the different categories of the fluid particles at $T = 2.18$ and $T = 6.43$, where $T = t\sqrt{g/H}$ is the non-dimensional time.

- *Additional diffusive terms due to the particle-shifting transport velocity*

The role of transport velocity of the particle-shifting in the governing equations (2) and the total energy of the system is negligible with invariant (or zero) potential energy (e.g., [25, 38, 49]). However, for long-term simulations and to capture the correct dynamics of violent free-surface flows (esp. with body forces), consideration of the deviation of velocity in the continuity and momentum equations becomes essential [5]. Here, in the context of the MPS method, we re-derive the additional terms (implemented by Sun et al. [5] in the SPH framework) to include the effects of the particle-shifting velocity in the governing equations. The particle shifting velocity, $\delta\hat{\mathbf{v}}_i$, is defined as the corrected particle shifting vector (Eq. (18)) divided by the time step, i.e.:

$$\delta\hat{\mathbf{v}}_i = {\delta\mathbf{r}_i}/{\Delta t}. \tag{29}$$

In Lagrangian derivatives of any arbitrary value (or vector), $f$, this velocity deviation modifying the transport velocity, $\mathbf{v} + \delta\hat{\mathbf{v}}$, appears as an extra term, $\nabla f . \delta\hat{\mathbf{v}}$ [5, 27]. Considering the particle number density and the momentum velocity, this term reads:

$$\begin{cases} \nabla n . \delta\hat{\mathbf{v}} = \nabla . (n\delta\hat{\mathbf{v}}) - n\nabla . (\delta\hat{\mathbf{v}}) \\ \nabla \mathbf{v} . \delta\hat{\mathbf{v}} = \nabla . (\mathbf{v} \otimes \delta\hat{\mathbf{v}}) - \mathbf{v}\nabla . (\delta\hat{\mathbf{v}}) \end{cases} \tag{30}$$

rewriting the governing equations of the EWC-MPS method (including the modified diffusive terms and the shear force) to:

$$\begin{cases} \dfrac{1}{n_i}\dfrac{Dn_i}{Dt} = -\langle\nabla.(\mathbf{v} + \delta\hat{\mathbf{v}})\rangle_i + \dfrac{1}{n_i}\langle\nabla.(n\delta\hat{\mathbf{v}})\rangle_i + D_i^m \\ \dfrac{D\mathbf{v}_i}{Dt} = -\dfrac{1}{\rho_i}\langle\nabla p\rangle_i + \mathbf{F}_i + \dfrac{1}{\rho_i}\langle\nabla.\boldsymbol{\tau}\rangle_i + \langle\nabla.(\mathbf{v}\otimes\delta\hat{\mathbf{v}})\rangle_i - \mathbf{v}_i\langle\nabla.(\delta\hat{\mathbf{v}})\rangle_i \\ \dfrac{D\mathbf{r}_i}{Dt} = \mathbf{v}_i + \delta\hat{\mathbf{v}}. \end{cases} \tag{31}$$

To estimate the additional terms, we should consider the mass and momentum conservation properties of the system [5]. The terms containing the velocity deviation are estimated by the following operators in the new system of equations:

$$\begin{cases} \dfrac{1}{n_i}\langle\nabla.(n\delta\hat{\mathbf{v}})\rangle_i = \dfrac{d}{n_0^2}\sum_{i\neq j}^{N}(n_j\delta\hat{\mathbf{v}}_j + n_i\delta\hat{\mathbf{v}}_i).\dfrac{\mathbf{e}_{ij}}{r_{ij}}W_{ij} \\ \langle\nabla.(\mathbf{v}\otimes\delta\hat{\mathbf{v}})\rangle_i = \dfrac{d}{n_0}\sum_{i\neq j}^{N}(\mathbf{v}_j\otimes\delta\hat{\mathbf{v}}_j + \mathbf{v}_i\otimes\delta\hat{\mathbf{v}}_i).\dfrac{\mathbf{e}_{ij}}{r_{ij}}W_{ij} \\ \mathbf{v}_i\langle\nabla.(\delta\hat{\mathbf{v}})\rangle_i = \mathbf{v}_i\dfrac{d}{n_0}\sum_{i\neq j}^{N}\dfrac{\delta\hat{\mathbf{v}}_j - \delta\hat{\mathbf{v}}_i}{r_{ij}}.\mathbf{e}_{ij}W_{ij}. \end{cases} \quad (32)$$

We approximate the first additional term inserted in the continuity equation, i.e., $\nabla.(n\delta\hat{\mathbf{v}})$, by the antisymmetric divergence operator and setting $1/n_i = 1/n_0$ to preserve the total mass as $\sum_i \dfrac{n_i}{n_0}\langle\nabla.(n\delta\hat{\mathbf{v}})\rangle_i V_i = 0$ [5]. Note that the transport velocity inside the divergence equation of the continuity equation contains the second additional term, $-n\nabla.(\delta\hat{\mathbf{v}})$. Similarly, for the momentum equation, we approximate the first additional term, $\nabla.(\mathbf{v}\otimes\delta\hat{\mathbf{v}})$, with the antisymmetric divergence in order to conserve the linear momentum of the system. Unlike the SPH model by Adami et al. [27] (further generalized by Zhang et al. [28]), we include the divergence of the velocity deviation (i.e., the second additional term, $-\mathbf{v}\nabla.(\delta\hat{\mathbf{v}})$) within the momentum equation. Although the approximated form of this additional term violates the momentum conservation law, it should be noted that the energy conservation of the system is not satisfied by the PS techniques [25, 37]. With the additional terms, the new formulation (31) aims to dissipate/eliminate the excessive energy inserted into the system by shifting the fluid particles.

Coupling the additional $\delta\hat{\mathbf{v}}$-terms in the governing equations with the new particle classification algorithm and the shifting vector (summarized in Figure 2), we introduce a new form of the CPS approach identified as CPS$^*$. This algorithm updates the velocity and particle number density fields by the discretized form of Eq. (31) after dispositioning the particles with $\delta\mathbf{r}_i$, given by Eq. (18) and (28). As originally pointed out by Jandaghian and Shakibaeinia [14], in problems with highly dynamic flow deformations, the fluid particles at the free-surface region are still vulnerable to particle-clustering, leading to unphysical fluid fragmentations (as their particle-shifting vector is canceled in the normal direction to the free surface). Similar to Jandaghian and Shakibaeinia [14], to improve the numerical stability of the free surface and the external particles, we append the DPC approach to the CPS$^*$ method. This hybrid technique, denoted as CPS$^*$+DPC, applies a corrected particle-shifting vector with the additional $\delta\hat{\mathbf{v}}$-terms over the fluid domain, while implementing the DPC method only among the free-surface, free-surface boundary, and splashed particles (i.e., $i,j \in \mathbb{F} \cup \mathbb{B}_F \cup \mathbb{E}$ in Eqs. (13) & (14)). It should be noted that the DPC would act as the only particle regularization technique of the particles with null particle-shifting in these regions (where $i \in \mathbb{E}$ or $(\alpha_i, \beta_i) = (0,0)$ by Eqs. (25) & (27)).

Figure 3 represents the numerical implementation of the new CPS*+DPC technique. We summarize the proposed enhancements of the CPS*+DPC scheme as follows:

- The new particle classification scheme:
    - recognizes the internal regions where unphysical volume expansion occurs ($i \in \mathbb{X}$),
    - identifies the fluid particles at the vicinity of the wall with the proposed non-local normal vectors ($i \in \mathbb{B}_w$),
    - detects the fluid particles at the intersections of the free-surface region and the vicinity of the wall ($i \in (\mathbb{F} \cup \mathbb{B}_F) \cap \mathbb{B}_w$), and
    - calculates the curvature of the interfaces by Eq. (26).
- The new particle-shifting formulation (Eq. (18)):
    - ignores the solid boundary particles in the shifting process of the fluid particles (Eq. (17)),
    - cancels the shifting vectors of the particles with the unphysical volume expansion ($i \in \mathbb{X} \to \delta \mathbf{r}_i = 0$),
    - through the binary multipliers, $\alpha_i$ and $\beta_i$, imposes some restricting conditions to the magnitude and/or direction of the particle shifting vector in the free-surface boundary and the vicinity of the wall, and the regions with extreme flow curvatures (Eqs. (24)-(27)), and
    - limits the magnitude of the shifting vector according to the particle's velocity and the spatial resolution (Eq. (28)).
- The additional diffusive terms due to the particle-shifting transport velocity (i.e., the $\delta\hat{\mathbf{v}}$-terms), which are re-derived in the context of the MPS method (Eq. (32)), update the velocity, position, and density of the particles based on the governing equations (31).
- The new DPC formulation, coupled with the CPS* algorithm, acts as the particle regularization technique of the particles within the free-surface regions and splashed particles (i.e., $i,j \in \mathbb{F} \cup \mathbb{B}_F \cup \mathbb{E}$) through equations (13) & (14).

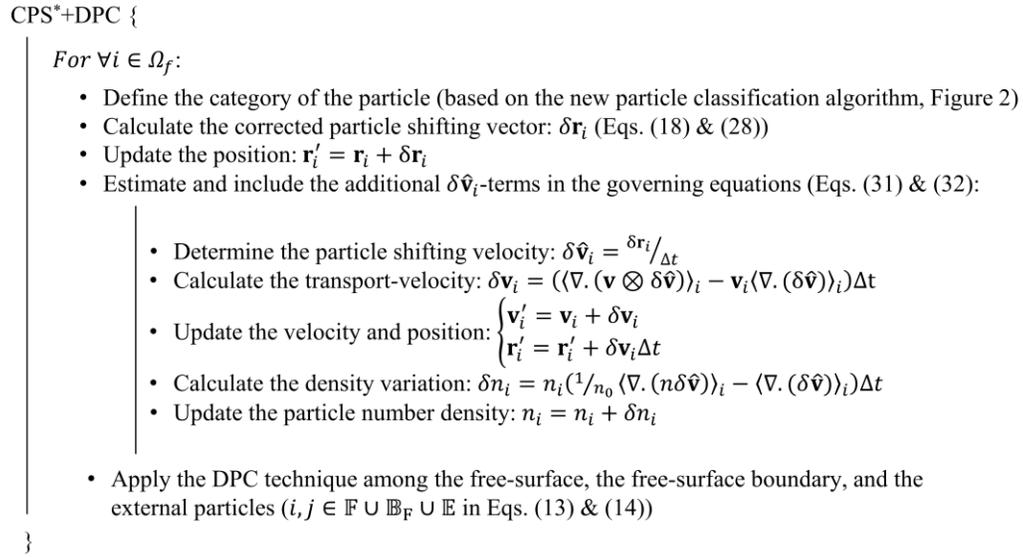

Figure 3- The sequential steps of the CPS$^*$+DPC algorithm as a hybrid particle regularization approach. The new particle classification algorithm is represented in Figure 2, and the corrected particle-shifting vector is given by Eqs. (18) & (28).

*3.3   Initial and solid boundary conditions*

Here, we apply the initial hydrostatic pressure to all the particles at $t = 0$ based on their initial position with respect to the free surface and considering the gravitational force as $g = 9.81 \ m/s^2$ in the negative y-direction. The inverse form of the EOS (i.e., $n_i^0 = n_0 (p_i^0/B_0 + 1)^{1/\gamma}$, in which $B_0 = c_0^2 \rho_0 /\gamma$), calculates the initial particle number density, $n_i^0$, which will be updated by the continuity equation.

The solid boundary particles consist of the wall and ghost particles (see [14]). The position of these particles remains fixed with respect to each other constructing the rigid walls. In this study, to simulate fluid-wall interactions, we adopt either the dynamic boundary condition [48] or the generalized wall boundary condition [50]. In these models, the pressure of the fluid particles approaching the wall increases through the continuity equation (i.e., in Eq. (4) considering the true velocity of the reservoir, $\mathbf{v}_w$, for the neighbor boundary particles). To implement the free-slip boundary condition for violent flow simulations, both models ignore the solid boundary particles in the fluid particle shear force calculations [30, 51].

Regarding the pressure of the wall particles, these boundary models implement different formulations. In the dynamic boundary condition, the continuity equation updates the density of the wall particles (with an adjusted wall velocity), and the equation of state calculates their pressure. Furthermore, the divergence of the velocity ignores the presence of ghost particles in the fluid and wall particle continuity equations. In the generalized wall boundary condition, the fluid particles' pressure at the vicinity of the wall is smoothed on the wall particles by including the gravity force and the wall's accelerations [50]. Then, the inverse form of the equation of state determines the wall's particle number density. However, in both methods, we directly assign the pressure of the nearest wall particle to the ghost particles.

*3.4 Time integration scheme*

We employ the second-order symplectic algorithm [52] for the temporal integration of the equations. This algorithm updates the interaction forces, the pressure field, and the particles' positions by dividing each time step, $\Delta t$, into two calculation stages with $\Delta t/2$ (see Appendix A). The viscosity of particles is recalculated at the beginning of each stage by estimating the gradient of velocity and the magnitude of the strain rate tensor. The adopted particle regularization techniques are implemented only in the second stage of the scheme before calculating the new particle number density and pressure fields (Figure A.1). Thus, to determine the transport velocity and regulate the particles' positions, $\Delta t$ is substituted by $\Delta t/2$ in Figure 3 and Eq. (14) for the CPS$^*$ and DPC techniques, respectively.

To determine the invariant time step of the integration scheme, $\Delta t$, we employ the Courant–Friedrichs–Lewy (CFL) stability condition as:

$$\Delta t = C_{CFL} \frac{l_0}{c_0} \tag{33}$$

in which $C_{CFL}$ is the CFL coefficient and the artificial sound speed, $c_0$, is a problem-dependent parameter determined based on the expected Mach number [9, 53]. One should note that in the problems studied as the violent and impact flows, even with the turbulence viscosity included in the shear stress calculations, the acoustic and the kinetic constraints remain dominant for determining the time step of the explicit model [53]. The solution algorithm is implemented on a parallel accelerated code run on the Graphical Processing Unit (GPU) using Compute Unified Device Architecture (CUDA) parallel programming language based on C++ (see Appendix A for more details).

*"Enhanced weakly-compressible MPS method for violent free-surface flows: Role of particle regularization techniques"*
*Jandaghian et al. (2021), ACCEPTED MANUSCRIPT, https://doi.org/10.1016/j.jcp.2021.110202*
## 4. Numerical results and discussions

In this study, we simulate challenging violent free-surface flow benchmark cases to evaluate the role of the proposed enhancements. The cases include the two-dimensional (2D) water dam-break and three-dimensional (3D) sloshing and obstacle impact problems.

*4.1  2D water dam-break*

This benchmark case has been the focus of numerous particle methods (e.g., [8, 9, 29, 54]). Capturing these complex flows, dominated by water-water and water-solid impacts, requires robust particle regularization techniques to eliminate possible numerical instabilities while conserving the total momentum and energy of the system [5].

We set up the initial configuration of the dam break based on the physical model by Lobovský et al. [55] (shown in Figure 4). With the initial height of the water column, $H$, and the initial particle distance, $l_0$, located on the Cartesian lattice, the spatial resolution is identified as $R = H/l_0$ (where in this study $R = $ 60, 100, and 200, which correspond to 11,840, 27,696, and 95,312 total numbers of particles, respectively). We simulate the time evolution of the dam break for 10 seconds and define the non-dimensional time as $T = t\sqrt{g/H}$. Based on the kinetic energy pressure constraint (for specifying the artificial sound speed of the weakly compressible model) and the maximum expected velocity, which is given as $\|\mathbf{v}\|_{max} = 2.0\sqrt{gH}$, we set the sound speed to 100 $m/s$, limiting the Mach number to 0.049 (see [53]). In all the cases, the $C_{CFL}$ coefficient is equal to 0.5, except in the model without the regularization technique, in which $C_{CFL} = 0.25$. We employ the dynamic boundary condition to simulate the interaction forces between the fluid and the rigid walls. In the DPC scheme, the maximum and minimum pressures of Eq. (15) are given as $2.3\rho g H$ (based on the analytical solution discussed in [53]) and $\rho g l_0$, respectively. In the case of the standard PC method being implemented, we set the constants of Eq. (12), $\varepsilon = 0.5$ and $\theta = 0.9$. In this problem, we normalize the global energies ($E_p$: the potential energy, $E_k$: the kinetic energy, $E_m$: the mechanical energy, and $Q_{PC}$: the total kinetic energy dissipated by the DPC technique, for which the formulations are given in Appendix B) by $\Delta E_m^{Fin} = E_m^\infty - E_m^0$, in which $E_m^0$ is the initial mechanical energy (equal to $E_p^0$) and $E_m^\infty$ is the final mechanical energy of the system (equal to the final expected potential energy with the water phase filling the width of the tank and reaching the equilibrium state).

26© 2021. This manuscript version is made available under the CC-BY-NC-ND 4.0
license https://creativecommons.org/licenses/by-nc-nd/4.0/

Concerning the short-term evolution of the flow ($T < 8$), we first validate the numerical (i.e., from the models with either the DPC or the CPS$^*$+DPC techniques or with no regularization technique) versus the experimental results of [55]. Figure 5 compares the time history of the local pressure signal at point *S1* and the propagation of the wave on the horizontal bed. To measure the local pressure, the pressure of the fluid particles inside the support radius, set to $1.6l_0$ from the sensor, are linearly averaged every 0.004 s without kernel-smoothing or any data-filtering (i.e., $p_{S1} = \sum_1^{N_s} p_j / N_s$, as $N_s$ is the number of fluid particles, $j$, where $\|\mathbf{r}_j - \mathbf{r}_{S1}\| \leq 1.6l_0$). For both parameters, the results of the simulations followed the same trends as the experimental measurements. Incompatibilities between the experimental and numerical results originate from the weak compressibility of the flow, ignoring the air phase, and the implemented solid boundary model. The numerical stability of the test case without the regularization technique can be evaluated from Figure 6. In this model, the particle-clustering becomes dominant with the progress of the impact events (i.e., as the wave impacts the front wall at $T \simeq 2.5$, a plunging jet forms and impinges the free surface at $T \simeq 6.06$). Numerical instability due to extreme inter-particle penetration terminated the simulation at $T \simeq 9.0$ (as the unphysical inter-particle forces ejected the clustered particles outside the predefined computational domain). Furthermore, as shown in the zoomed-in sections of Figure 6, unphysical fluid fragmentations exist, especially in the flow regions with highly dynamic deformations. Eliminating these numerical issues, which hinder long-term simulations of violent flows and affect flow evolutions, requires a great emphasis on the adoption of rigorous particle regularization techniques.

In this part, we investigate the role of the proposed DPC regarding the stability and energy evolution of the system. Figure 7 depicts that both the DPC and PC techniques guarantee the long-term simulation of the dam break by resolving the numerical stabilities. However, the model with the PC method still suffers from noisy pressure fields, which may lead to inaccurate flow predictions. In contrast, the DPC technique effectively reduces the high-frequency pressure fluctuations (specifically illustrated in the zoomed-in sections of Figure 7). Regarding the energy variations of the fluid phase, Figure 8 shows that both the PC and DPC methods reached the final expected equilibrium state (i.e., $\Delta E_p/\Delta E_m^{Fin} = -1$) without any manipulation of the potential energy. The kinetic, potential, and mechanical energies of these models followed almost identical patterns during the impact events (between $T = 2.5$ and 17.5), since their formulations did not affect the total linear momentum of the system. Moreover, in comparison to the model without regularization (which is stable till $T \simeq 9.0$), both the PC and DPC techniques represent similar energy profile evolutions. However, the DPC dissipates the kinetic energy almost two times as much as the standard PC technique. This confirms that the DPC ensures numerical stability without changing the overall energy of the system, while acts as an efficient artificial viscosity formulated by the pairwise and dynamic repulsive and collision terms. Furthermore, Figure 8 depicts that with the higher spatial resolutions ($R = 60, 100$, and $200$), the heat produced by the DPC, i.e., $Q_{PC}$, reduces which confirms the numerical convergence of the model (especially during the violent dam break flow).

In the next step, we study the effectiveness of the corrected particle-shifting algorithms in capturing the flow and energy evolutions. As shown in the right column of Figure 9, the CPS$^*$+DPC algorithm simulates the evolution of the dam break without unphysical volume expansion; however, the same model without the additional $\delta \mathbf{v}$-terms expands the volume of the fluid phase, particularly when $T \gtrsim 8.0$ (Figure 9, left column). To further clarify the role of the $\delta \hat{\mathbf{v}}$-terms, Figure 10 compares the energy profiles of the CPS$^*$+DPC method (with and without the $\delta \hat{\mathbf{v}}$-terms) versus the original CPS+PC [14]. It shows that the continuous application of particle-shifting with the original CPS+PC increases the potential energy of the system considerably after $T \gtrsim 8.0$. Even limiting the magnitude of the shifting vector (by Eq. (28)) in the original CPS+PC does not reduce the accumulation of errors, which leads to inaccurate estimation of the mechanical energy in a long-term simulation. The new CPS scheme without the $\delta \hat{\mathbf{v}}$-terms decreases these errors by canceling the shifting vectors of the expanded particles (i.e., $i \in \mathbb{X} \to \delta \mathbf{r}_i = 0$). The consistent CPS$^*$+DPC method completely dissipates the excess energy inserted into the system noting that the potential energy converges to the final expected value (i.e., no volume expansion occurs). Moreover, by increasing the spatial resolution, the CPS$^*$+DPC formulations represent slightly fewer errors in predicting the final potential energy. Nevertheless, as shown in Figure 10, the kinetic energy is noticeably dissipated with the CPS$^*$+DPC technique (between $T = 6.0$ and $17.5$). This shows that although implementation of the $\delta \hat{\mathbf{v}}$-terms is required for long-term simulations (to avoid unphysical volume expansion), these terms can affect the kinetic energy and therefore the mechanical behavior of the 2D dam-break flow during impact events. We should also note that the additional terms might only be essential for the weakly compressible models and can be neglected for a fully incompressible model (subjected to further investigations).

Figure 11 demonstrates that the EWC-MPS with the proposed stabilization techniques ensures the long-term-simulations of these complex flows and represents noise-free pressure fields. However, we observe that the CPS$^*$+DPC method slightly alters the flow evolution (particularly at $T = 8.98$) by excessive dissipation of the kinetic and potential energies (also shown in Figure 10). Meanwhile, the model with the DPC technique predicts the flow and pressure evolutions without the numerical instabilities or affecting the evolutions of the global energies. One should also note that in the CPS$^*$+DPC technique although the PS vector is neglected for the splashed particles and where $(\alpha_i, \beta_i) = (0,0)$ (through Eqs. (25) & (27)), the DPC technique is effectively ensuring the numerical stability in these regions (e.g., see the corners of the fluid domain of the dam-break problem in Figure 2).

To further compare the numerical efficiency of the two techniques, the calculation time per iteration of simulations, i.e., $t^{iter.}$ =Total calculation time/Total number of time steps, are represented in Table 1 for the three spatial resolutions (where all the simulations use an NVIDIA V100 Volta GPU device for solving the main temporal loop of the algorithm implemented with CUDA C++ parallel programming). The comparison shows that the 2D model with the DPC technique is ~18% faster than the simulations with the CPS$^*$+DPC algorithm. Regardless of the coding implementations and the used computer hardware, this result was expected as the DPC technique does not require free surface detections and is simple to implement; while the CPS$^*$+DPC technique, with the particle classification algorithm and approximating the additional terms, introduces several conditional and arithmetic operations in the model that would increment the calculation costs.

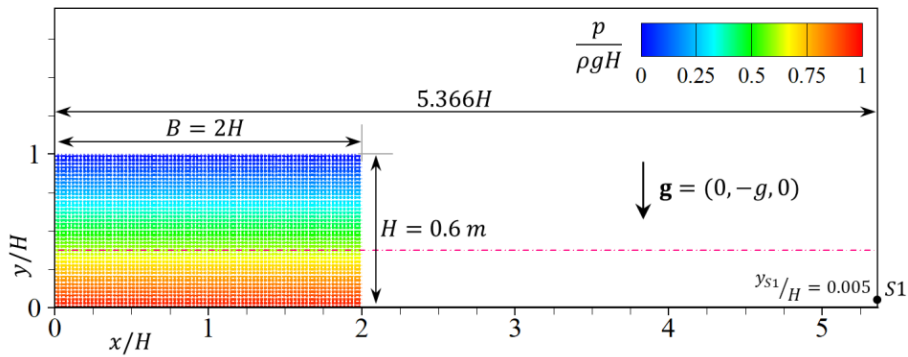

Figure 4- Initial configuration and pressure contour of the 2D water dam-break based on the experimental setup by Lobovsky et al. [55]. The dash-dot line is the expected free surface of the fluid flow filling the width of the tank at the final equilibrium state.

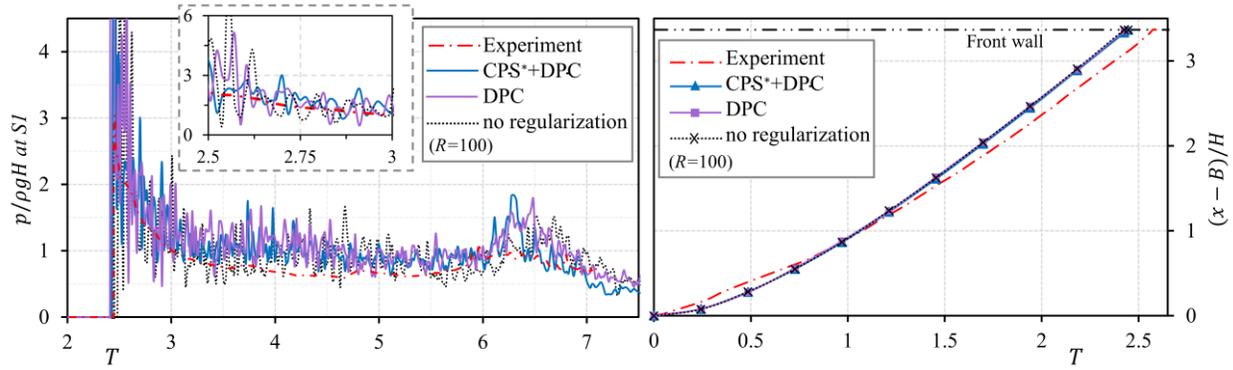

Figure 5- Numerical results of the models ($R = H/l_0 = 100$) with the CPS$^*$+DPC, DPC, and without regularization techniques versus the experimental results of Lobovský et al. [55]: the time history of averaged local water pressure, $p$, at $S1$ (left) and the wave propagation (right).

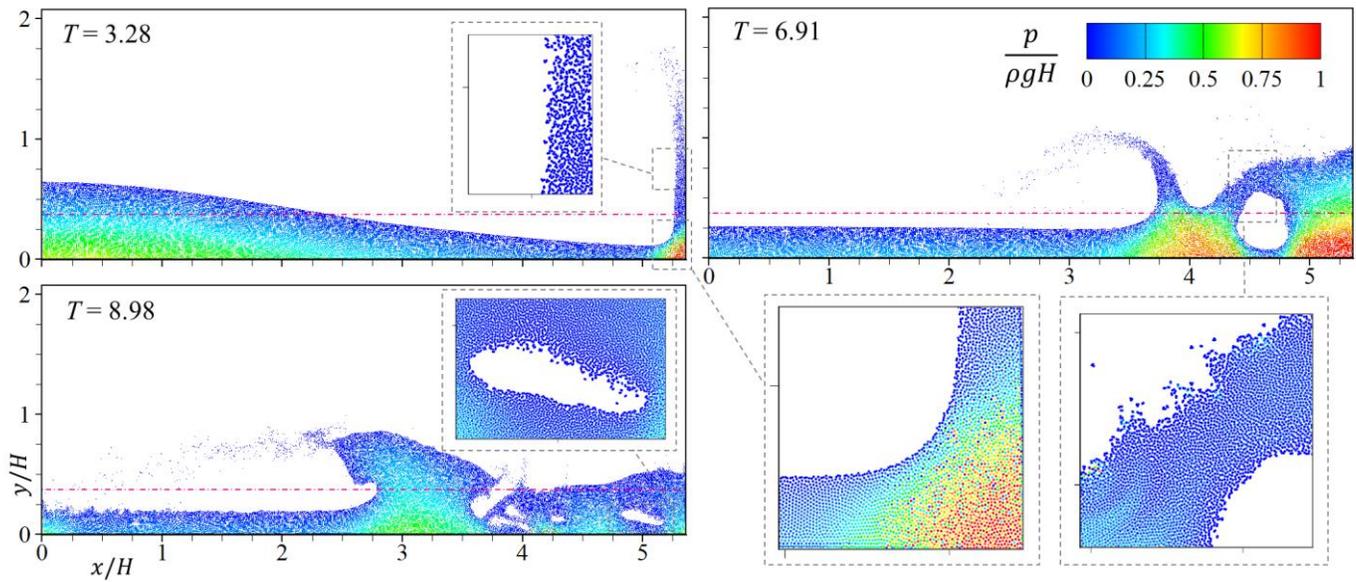

Figure 6- The dam break problem without the particle regularization techniques ($R = 200$). The dash-dot line is the expected free surface of the fluid flow filling the width of the tank at the final equilibrium state.

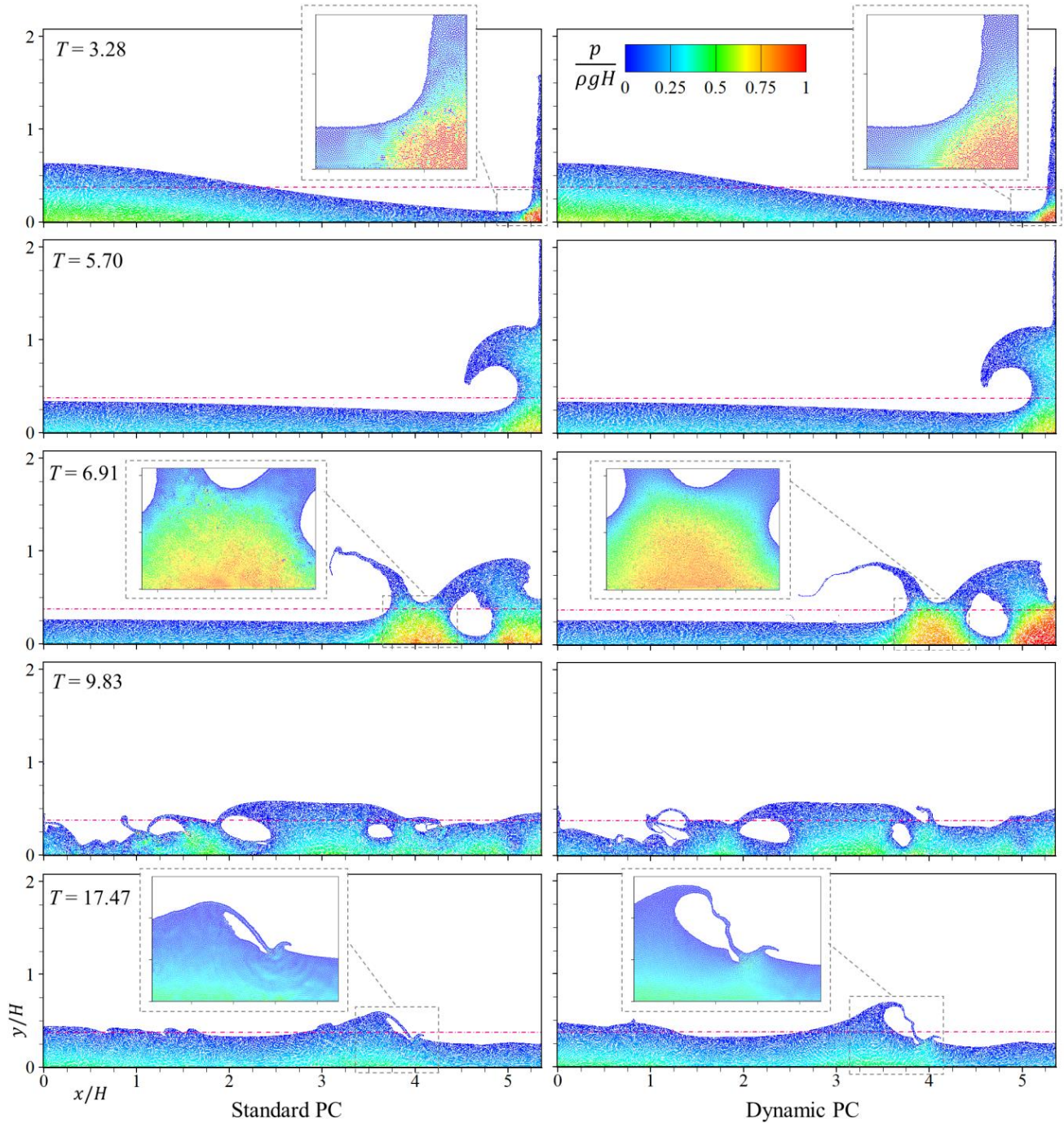

Figure 7- Time evolution of the dam break problem with the standard PC (left column) and the DPC (right column) techniques ($R = 200$). The color contour represents the non-dimensional pressure field at $T = t\sqrt{g/H} = 3.28, 5.70, 6.91, 9.83$, and $17.47$ (from the top row to the bottom row, respectively). The dash-dot line is the expected free surface of the fluid flow filling the width of the tank at the final equilibrium state.

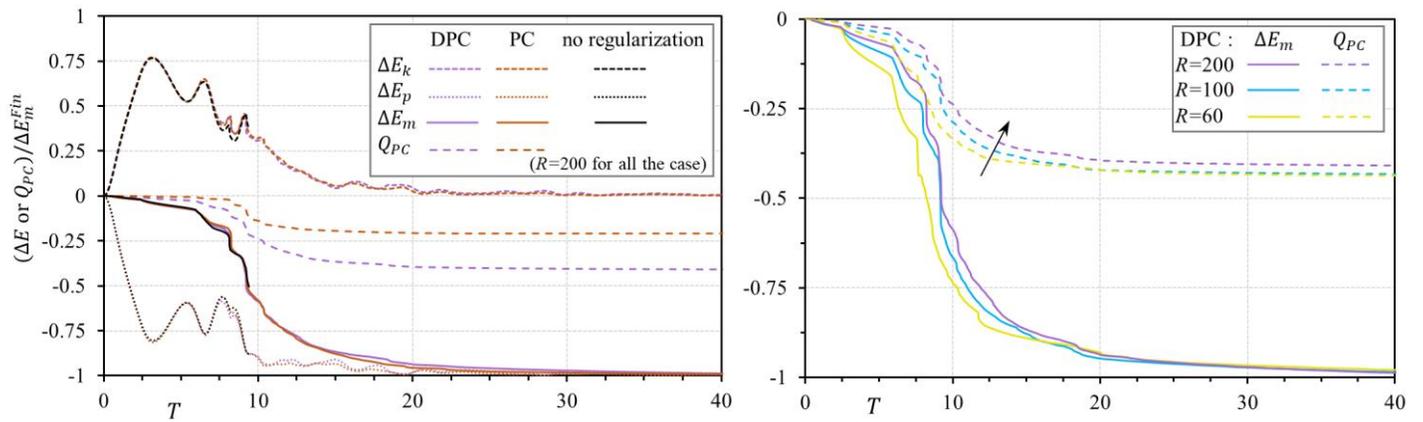

Figure 8- Time history of the kinetic, $\Delta E_k$, potential, $\Delta E_p$, and mechanical, $\Delta E_m$, energies of the dam break test case with the DPC vs PC methods (left graph). The total kinetic energy dissipation by the PC techniques (accumulated over time) is denoted as $Q_{PC}$. The right graph represents the particle convergence of the 2D dam-break simulation as the total energy dissipation reduces with increasing the spatial resolution, $R = H/l_0$.

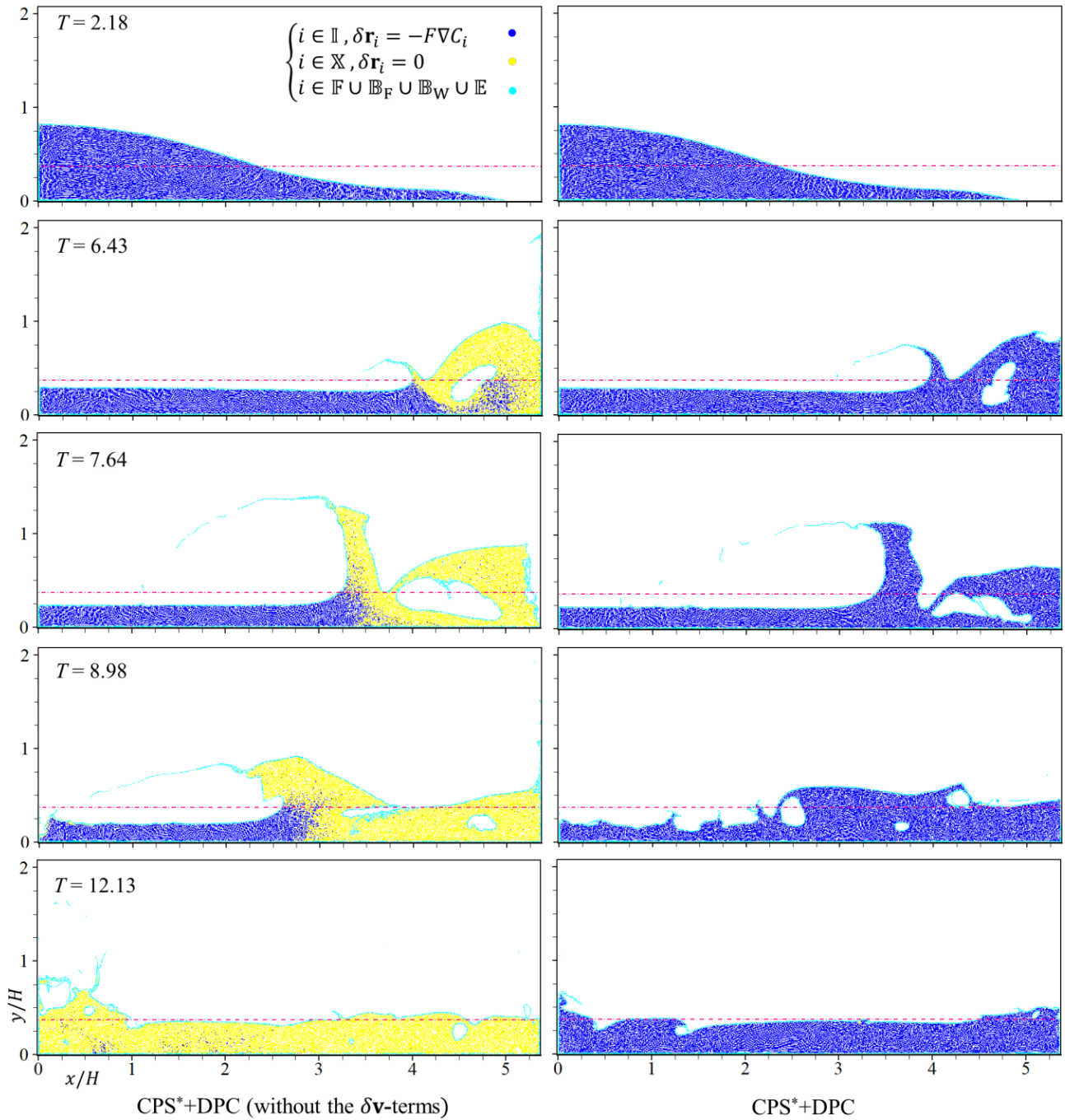

Figure 9- Time evolution of the dam break problem simulated by the CPS*+DPC technique with and without the $\delta\hat{\mathbf{v}}$-terms (represented in the right and left columns, respectively). The resolution is set to $R = 200$ and the expanded regions, i.e., $i \in \mathbb{X}$, are identified with the yellow particles. The dash-dot line is the expected free surface of the fluid flow filling the width of the tank at the final equilibrium state.

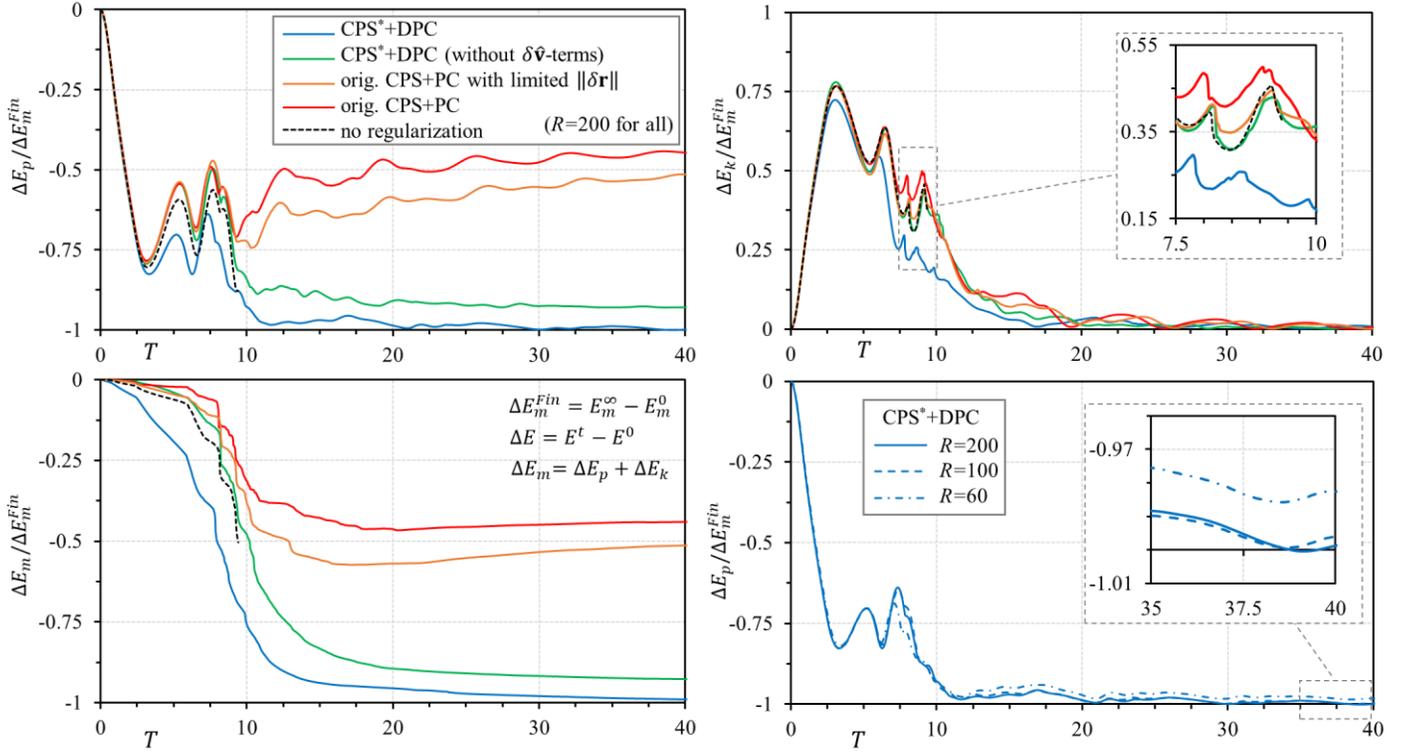

Figure 10- The evolution of energy components ($E_p$: the potential energy (top left), $E_k$: the kinetic energy (top right), and $E_m$: the mechanical energy (bottom left) (defined by Eq. (B.1)) of the 2D dam-break problem with different CPS techniques and resolutions. The bottom right graph shows the convergence of the potential energy to the expected final state (i.e., $\Delta E_p/\Delta E_m^{Fin} = -1$) by increasing the spatial resolution, $R = H/l_0$.

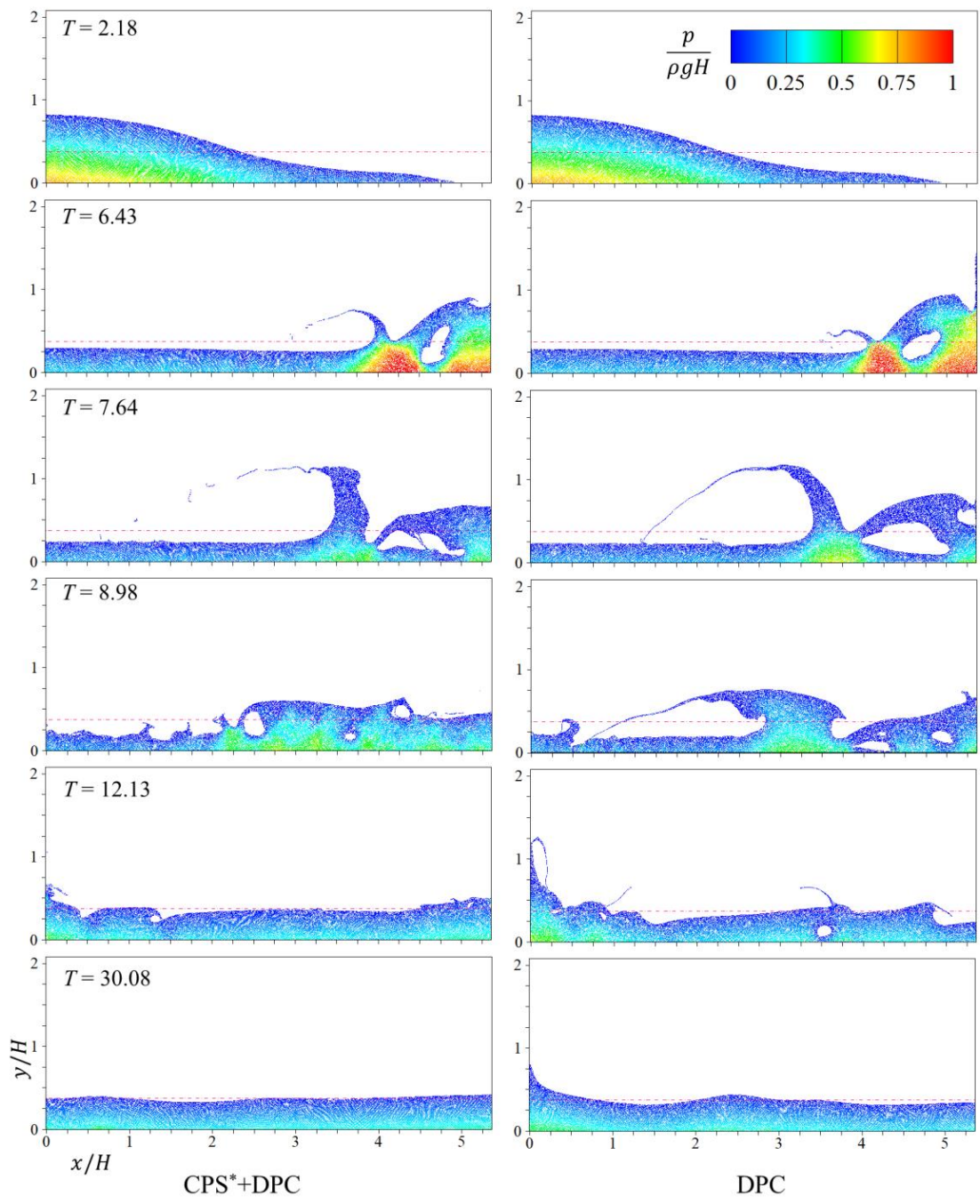

Figure 11- Time evolution of the dam break problem with the CPS$^*$+DPC (left column) and the DPC (right column) techniques ($R = 200$). The color contour represents the non-dimensional pressure field. The dash-dot line is the expected free surface of the fluid flow filling the width of the tank at the final equilibrium state.

Table 1- Calculation time per iteration (i.e., $t^{iter.}$ =Total calculation time/Total number of time steps) of the 2D water dam-break problem with different spatial resolutions, $R$. The last row of the table represents the rate of increase in the calculation

time obtained by the CPS$^*$+DPC versus the DPC technique. The GPU accelerated code is run on an NVIDIA V100 Volta GPU device.

| $R = H/l_0$ | | 60 | 100 | 200 |
|---|---|---|---|---|
| Number of fluid particles, $\Omega_f$ | | 7,200 | 20,000 | 80,000 |
| Calculation time per iteration, $t^{iter.}$ (sec.) | DPC | 7.42E-03 | 8.49E-03 | 2.04E-02 |
| | CPS$^*$+DPC | 8.69E-03 | 1.01E-02 | 2.41E-02 |
| $[t^{iter.}(\text{CPS}^*+\text{DPC})/t^{iter.}(\text{DPC})-1]\times 100$ | | 17.11 | 18.96 | 18.13 |

## *4.2  3D water sloshing in a rectangular reservoir*

Here we simulate 3D water sloshing in a rectangular reservoir that moves with a harmonic motion in the longitudinal direction. Under the gravitational force and with this motion's frequency being close to the natural frequency of water, a violent flow consisting of bore breaks and impact events forms within the tank [39]. This highly non-linear free-surface flow has been modeled by particle methods to show their effectiveness in dealing with such complex features (e.g., [21, 29, 39, 56]).

We set up the numerical model based on the physical model by Rafiee et al. [56]. A horizontal channel with dimensions $L \times D \times W$ (in the x-, y-, and z-directions, where $L = 1.3$, $D = 0.9$, and $W = 0.1\,m$) is partially filled with water with an initial height of $H = 0.2D$ (Figure 12). To validate the results with the experimental measurements, we extract the local pressure (every 0.004 seconds) on the side-wall of the channel at points $P1$, $P2$, and $P3$ as identified in Figure 12 (with the same formulation used in the 2D dam-break test case). The numerical results are represented for 8 seconds of the simulations. The frequency, $f$, and amplitude, $A_0$, of the harmonic motion are set to $0.496\,s^{-1}$ and $0.1\,m$, respectively. Sinusoidal excitation updates the positions and velocities of the solid boundary particles ($i \in \Omega_s$) in the x-direction as $\mathbf{r}_i = \mathbf{r}_i^0 + \{A_0 \sin(2\pi f_0 t), 0, 0\}$ and $\mathbf{v}_i = \{2\pi A_0 f_0 \cos(2\pi f_0 t), 0, 0\}$ at the midpoint and the end of each time step. Furthermore, we assign the initial hydrostatic pressure and the velocity as equal to $\mathbf{v}_i^0 = \{2\pi A_0 f_0, 0, 0\}$ for all the particles at $t = 0$. In this problem, we implement the generalized wall boundary condition from Adami et al. [50], which includes the gravity force, **g**, and the acceleration of the rigid walls, $\mathbf{a}_{w_i} = \{-4A_0(\pi f_0)^2 \sin(2\pi f_0 t), 0, 0\}$, in the pressure calculation for the wall particles. We define the spatial resolution of the simulations as $R = L/l_0$ set to 100, 200 and 260, which corresponds to the total numbers of 90,972, 365,464 and 661,952 particles located on the cubic lattice. For all the test cases $C_{CFL} = 0.25$ and $c_0 = 50\,m/s$, the maximum expected velocity, given as $\|\mathbf{v}\|_{max} = 2.0\sqrt{gH}$, limits the Mach number to 0.053 [53]. The maximum and minimum pressures of the DPC technique (in Eq. (15)) are determined as 15 $kPa$ (according to the experimental measurements in [56]) and $\rho g l_0$, respectively.

Figure 13 plots the numerical local pressure at points $P1$, $P2$, and $P3$ versus the experimental results of [56]. The time evolutions of the pressure profiles with either the DPC or the CPS$^*$+DPC techniques are in good agreement with those from the experimental measurements. In the physical model, the air pockets entrapped as the flow impacts the wall avoid a sudden increase in local pressure; however, the single-phase numerical models, by ignoring the air phase, neglect the cushioning effects [29, 39, 56]. This issue results in a discrepancy between the numerical and experimental measurements, manifested as overestimations of the numerical pressure when the wave reaches the sidewall at $t = 2.5$, 4.5, and 6.5 seconds (Figure 13).

Figure 14 illustrates the flow evolution and pressure fields, comparing the simulations with the CPS$^*$+DPC and DPC where $R = 260$. Both models show identical flow evolutions while ensuring long-term numerical stability by resolving the particle-clustering problem. Moreover, with the modified diffusive term, the simulations represent smooth pressure fields. In Figure 15, the new CPS$^*$+DPC technique with the implemented $\delta\hat{\mathbf{v}}$-terms preserves the overall volume of the fluid phase (shown in the right column); nevertheless, the same model without the additional $\delta\hat{\mathbf{v}}$-terms leads to unphysical volume expansion after a few cycles (see left column). Figure 16 represents the new dynamic particle classification in the 3D configuration of the fluid flow. With this algorithm, the CPS$^*$+DPC technique deals dynamically with the extreme flow curvatures (including the corners) and the intersection of the particles detected as the free-surface boundary and the wall boundary particles.

The temporal evolutions of the global energies of the system are illustrated in Figure 17, where the non-dimensional time is given as $T = t\sqrt{g/L}$. Here, the components of energy, i.e., $\Delta E$ and $Q_{PC}$ (see Appendix B), are normalized by the initial mechanical energy, $E_m^0 = E_p^0 + E_k^0$. We observe that CPS$^*$+DPC without the $\delta\hat{\mathbf{v}}$-terms increases the potential energy over time (also shown in the left column of Figure 15). On the other hand, implementing the $\delta\hat{\mathbf{v}}$-terms dissipates the excessive potential energy and predicts less kinetic energy versus the DPC technique. Furthermore, the results show that with a higher spatial resolution ($R = 200$ and 260), the heat produced by the DPC technique, i.e., $Q_{PC}$, decreases.

Moreover, Table 2 represents the calculation time per iteration ($t^{iter.}$) of the 3D simulations with the three spatial resolutions using an NVIDIA V100 Volta GPU device for solving the main temporal loop of the algorithm. Comparing the calculation time per iteration demonstrates that the CPS$^*$+DPC algorithm increases the calculation cost by ~20%. This result highlights that the DPC technique while being as effective as the CPS$^*$+DPC technique with less energy dissipation is simple to numerically implement since does not involve the time-consuming conditional operations of the CPS$^*$+DPC algorithm (required for the particle classifications and approximating the additional terms).

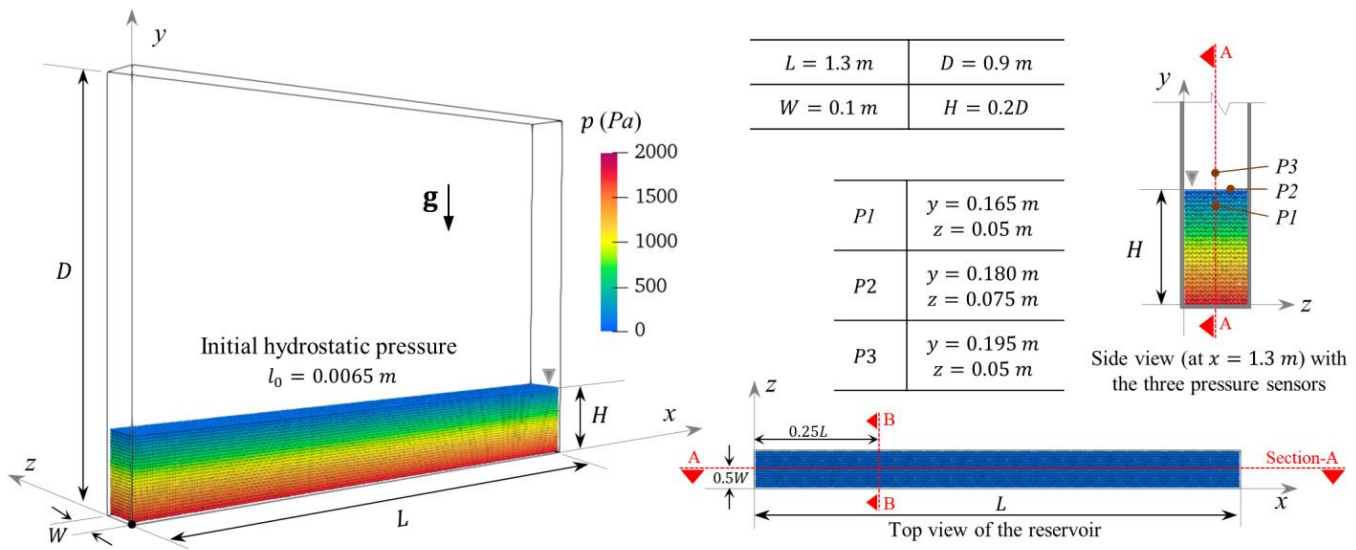

Figure 12- Initial configuration and pressure field of the 3D water sloshing problem. The locations of the three pressure sensors (*P1*, *P2*, and *P3*, which in the physical model by Rafiee et al. [56] correspond to sensors *P180*, *P078*, and *P136*, respectively) are specified on the side view of the reservoir. Section A-A and Section B-B, identified on the top view of the reservoir, are used to represent the numerical results.

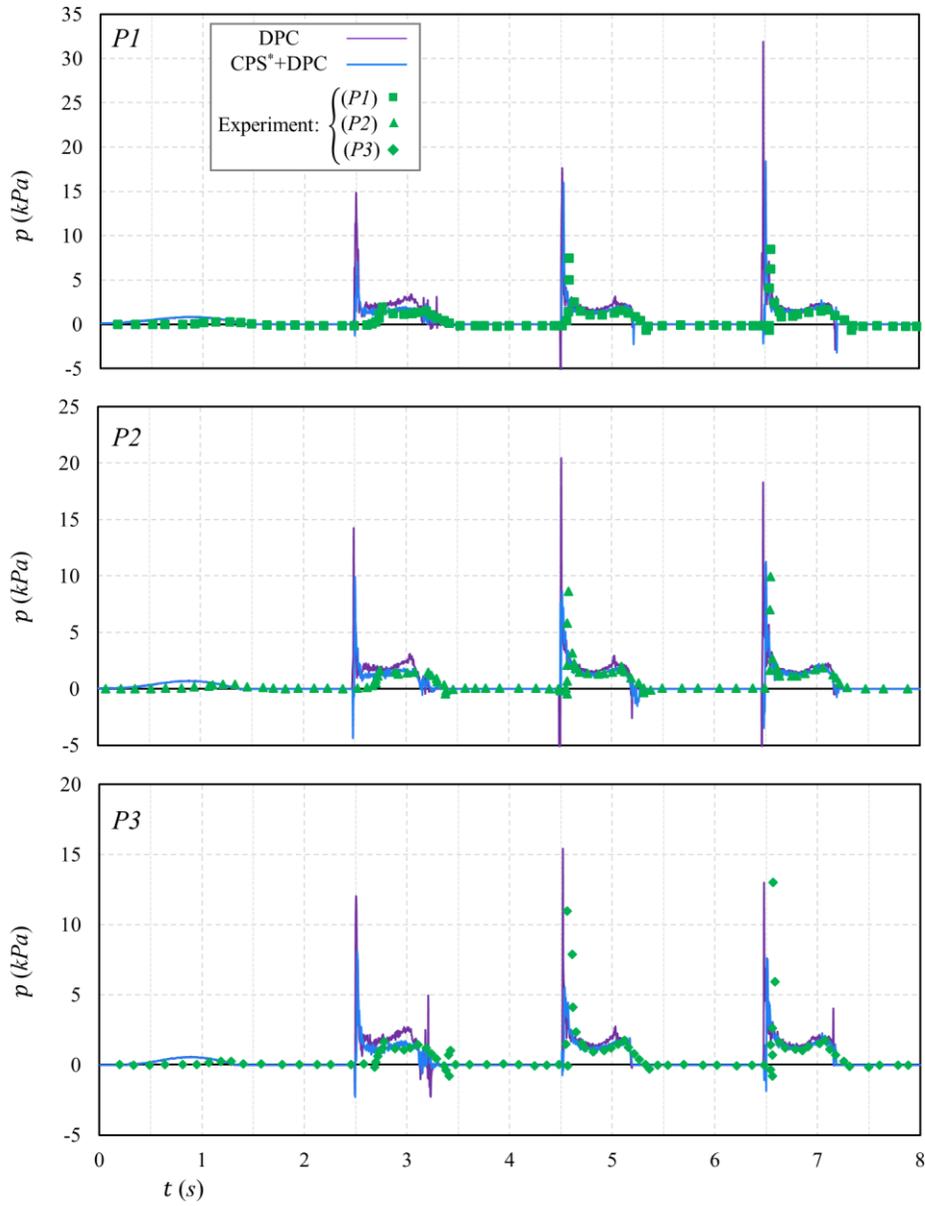

Figure 13- Local numerical pressure (with $R = L/l_0 = 200$) versus the experimental measurements of Rafiee et al. [56] at the $P1$ (top), $P2$ (middle), and $P3$ (bottom) points.

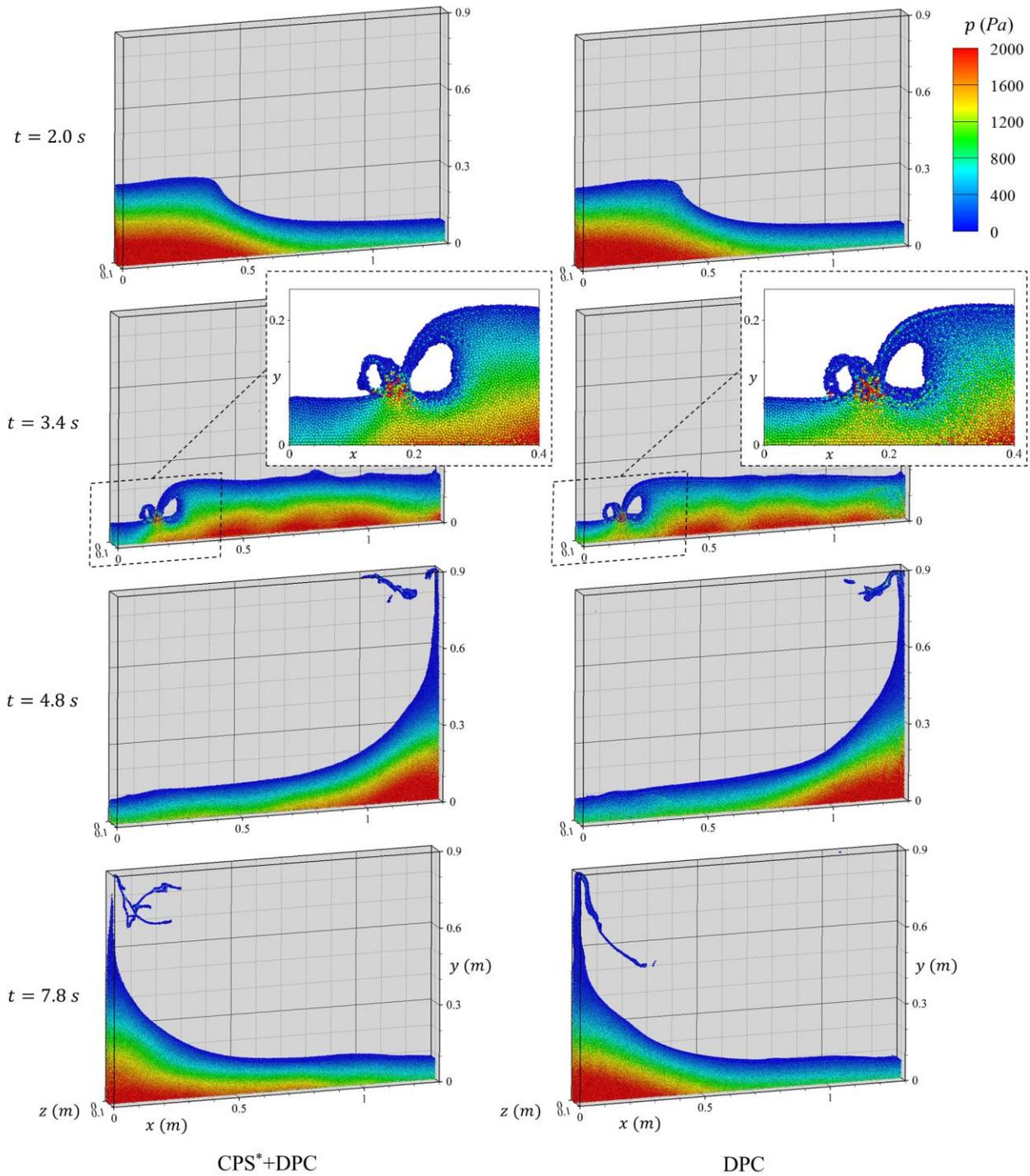

Figure 14- Flow evolutions and the pressure field of the water sloshing problem (at Section A-A) simulated by the model with the CPS*+DPC (left column) and the DPC (right column) techniques, where $R = 260$.

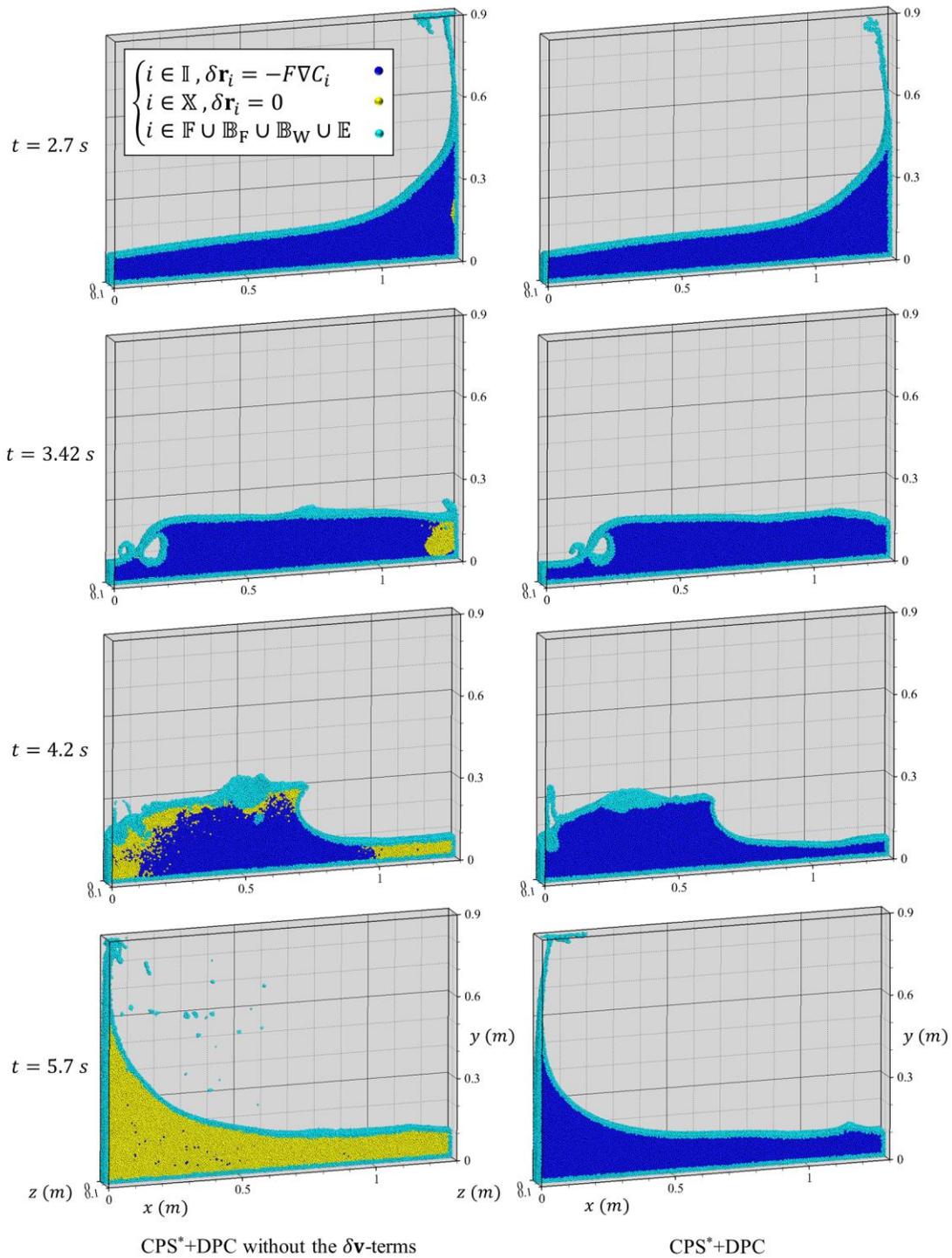

CPS*+DPC without the $\delta\mathbf{v}$-terms  CPS*+DPC

Figure 15- Time evolution of the 3D water sloshing problem (at Section A-A) using the CPS*+DPC techniques with and without the additional $\delta\hat{\mathbf{v}}$-terms (represented in the right and the left columns, respectively). The resolution is set to $R = 200$, and the expanded regions, i.e., $i \in \mathbb{X}$, are identified with the yellow particles.

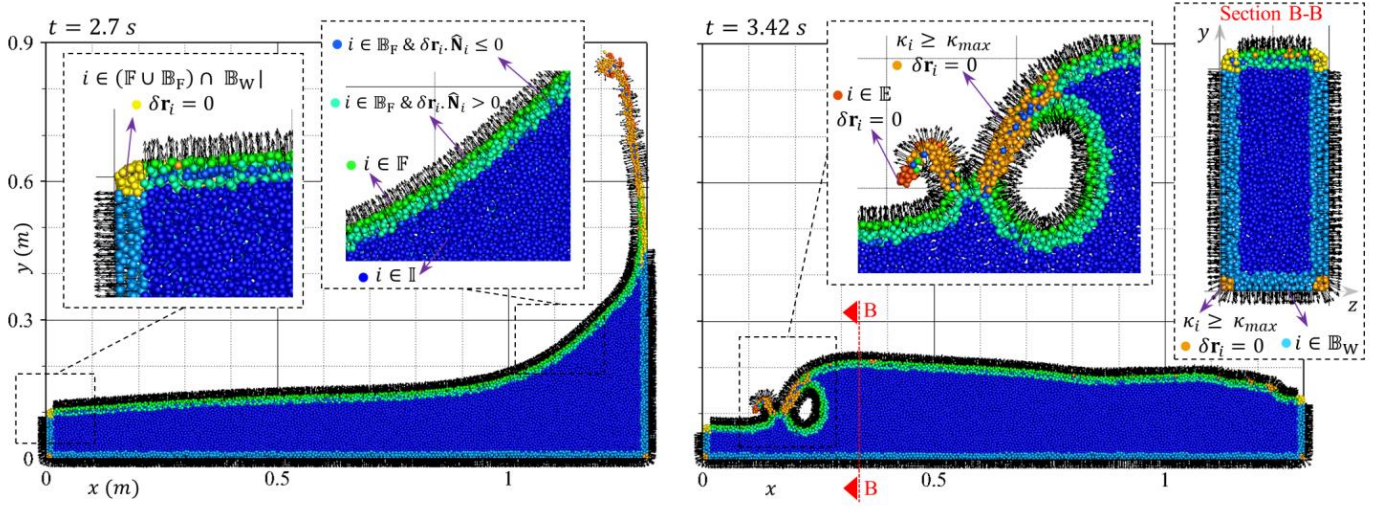

Figure 16- Particle classification and normal vectors of the 3D sloshing problem at Section A-A and Section B-B with the CPS$^*$+DPC technique and $R = 200$.

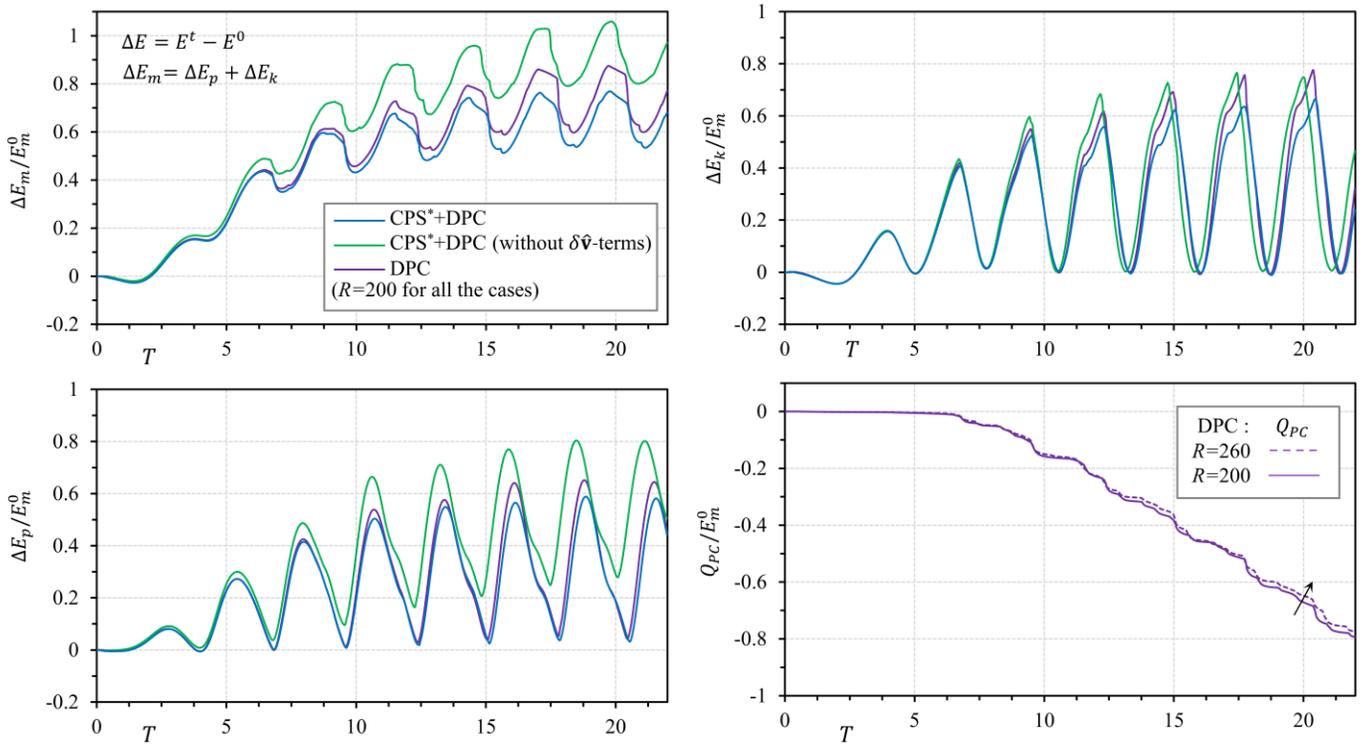

Figure 17- Evolution of energy components ($E_m$: the mechanical energy (top left), $E_k$: the kinetic energy (top right), $E_p$: the potential energy (bottom left), and $Q_{PC}$: the total kinetic energy dissipated by the DPC technique (bottom right) (defined by Eqs. (B.1) & (B.2)) of the 3D water sloshing test case with the DPC and CPS$^*$+DPC (with and without the $\delta\hat{\mathbf{v}}$-terms) techniques, where $R = 200$. The non-dimensional time, $T$, is given as $t\sqrt{g/L}$, and the energy variations, $\Delta E = E^t - E^0$, are normalized by the initial mechanical energy of the system, $E_m^0 = E_p^0 + E_k^0$.

Table 2- Calculation time per iteration (i.e., $t^{iter.}$ =Total calculation time/Total number of time steps) of the 3D water sloshing problem with different spatial resolutions, $R$. The last row of the table represents the rate of increase of the calculation time with the CPS$^*$+DPC versus the DPC technique. The GPU accelerated code is run on an NVIDIA V100 Volta GPU device.

| $R = L/l_0$ | | 100 | 200 | 260 |
|---|---|---|---|---|
| Number of fluid particles, $\Omega_f$ | | 11,700 | 81,000 | 187,200 |
| Calculation time per iteration, $t^{iter.}$ (sec.) | DPC | 5.83E-02 | 1.33E-01 | 3.67E-01 |
| | CPS$^*$+DPC | 6.68E-02 | 1.66E-01 | 4.47E-01 |
| $[t^{iter.}$(CPS$^*$+DPC)$/t^{iter.}$ (DPC)$-1]\times 100$ | | 14.58 | 24.81 | 21.79 |

*4.3  3D water dam-break against a rigid obstacle*

In this test case, under the gravitational force, a column of water collapses and impacts a rigid cuboid obstacle. Particle methods have widely simulated this problem as a benchmark case for studying violent flows and fluid-structure interactions (e.g., [30, 36, 39, 57-59]).

We employ the experimental model from Kleefsman et al. [60] to set up and validate the numerical simulations. Figure 18 illustrates the initial configuration of the problem, the location of points $P1$ and $P2$, and the vertical lines $H1$ and $H2$ used for probing the local pressure and the depth of the water, respectively. With the spatial resolution given as $R = H/l_0 = 55$, the simulation contains 1,044,444 particles, including the fluid and solid boundary particles located on the cubical lattice. In this benchmark case, the generalized wall boundary model from Adami et al. [50] simulates the interaction forces between the fluid and the rigid bodies. The CFL coefficient and the Mach number are equal to 0.5 and 0.046, respectively, with the sound speed set to 100 $m/s$ and $\|\mathbf{v}\|_{max} = 2.0\sqrt{gH}$. Like the dam-break test case, the maximum and minimum background pressures of the DPC formulation are determined as 2.3$\rho g H$ and $\rho g l_0$, respectively.

The numerical results of the proposed model with either the DPC or CPS$^*$+DPC represent trends almost similar to the experimental measurements (Figure 19). The depth of water at the $H1$ and $H2$ lines (extracted every 0.05 seconds for three spatial resolutions, $R = 13.75, 27.5,$ and $55$) shows that the flow evolutions are comparable with the experimental profiles. However, the absence of the air phase and its cushioning effects in the numerical simulation introduces some discrepancies between the measurements, particularly during the main impact events (i.e., $1.75 \lesssim T = t\sqrt{g/H} \lesssim 8.5$). This issue results in oscillatory impact pressures at points $P1$ and $P2$ (extracted every 0.004 seconds) and leads to a delay in

*"Enhanced weakly-compressible MPS method for violent free-surface flows: Role of particle regularization techniques"*
*Jandaghian et al. (2021), ACCEPTED MANUSCRIPT, https://doi.org/10.1016/j.jcp.2021.110202*the second wave that impacts the obstacle at $T \approx 20.0$. These incompatibilities between the numerical and experimental results are also noticeable in the measurements of other single-phase particle methods, e.g., [57, 58]. However, the results confirm that with increasing the spatial resolution the pressure fluctuations considerably decrease showing the numerical convergence of the 3D model (Figure 19).

Figure 20 compares the flow evolution with the DPC versus the CPS$^*$+PC techniques. We observe that the proposed EWC-MPS method, by resolving the numerical instabilities, can capture the complex flow deformations of this benchmark case. As the wave impacts at the rigid step, the flow splashes up and two jets form on the horizontal bed and impact the vertical wall of the reservoir; the flow, while overtopping the obstacle, collides with the returning jets and forms plunging waves [36, 39]. Furthermore, Figure 21 shows that the diffusive model represents regular pressure fields at Section A-A during the impact events. The evolution of the energy components of the system, normalized by $\Delta E_m^{Fin} = E_m^\infty - E_m^0$, is plotted in Figure 22. Both particle regularization techniques introduce less energy dissipation by increasing the spatial resolution, confirming the numerical convergence of the 3D simulations. Also, similar to the previous test case, the DPC technique is slightly less dissipative than the CPS$^*$+DPC formulations. Overall, the evolution of the global energies obtained by both techniques is almost identical (with $R = 55$), and they eventually reach the expected potential energy at $T \approx 25$ (i.e., $\Delta E_p/\Delta E_m^{Fin} = -1$).

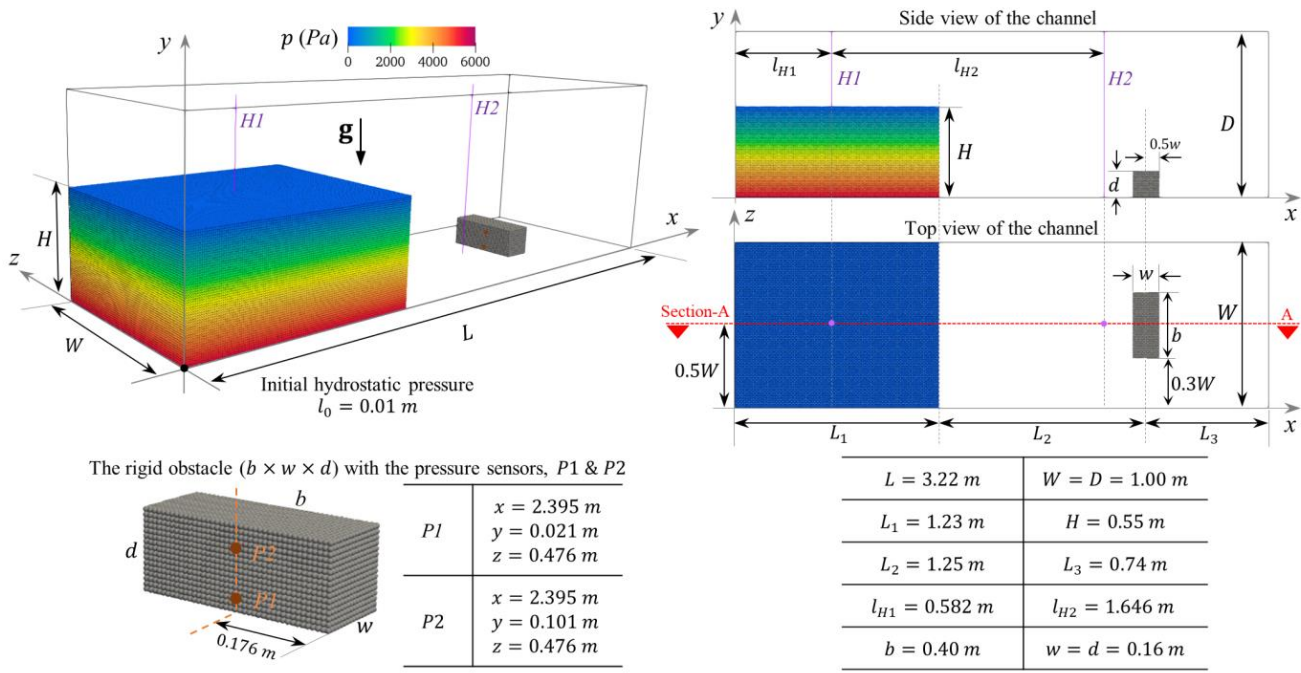

44© 2021. This manuscript version is made available under the CC-BY-NC-ND 4.0
license https://creativecommons.org/licenses/by-nc-nd/4.0/

Figure 18- Initial configuration of the 3D water dam-break against the rigid obstacle, based on the physical model by Kleefsman et al. [60]. The pressure sensors, $P1$ and $P2$, are located on the side wall of the obstacle, and the vertical water heights in the y-direction at $H1$ and $H2$ are probed. Section A-A, shown on the top view of the channel, is used to represent the numerical results.

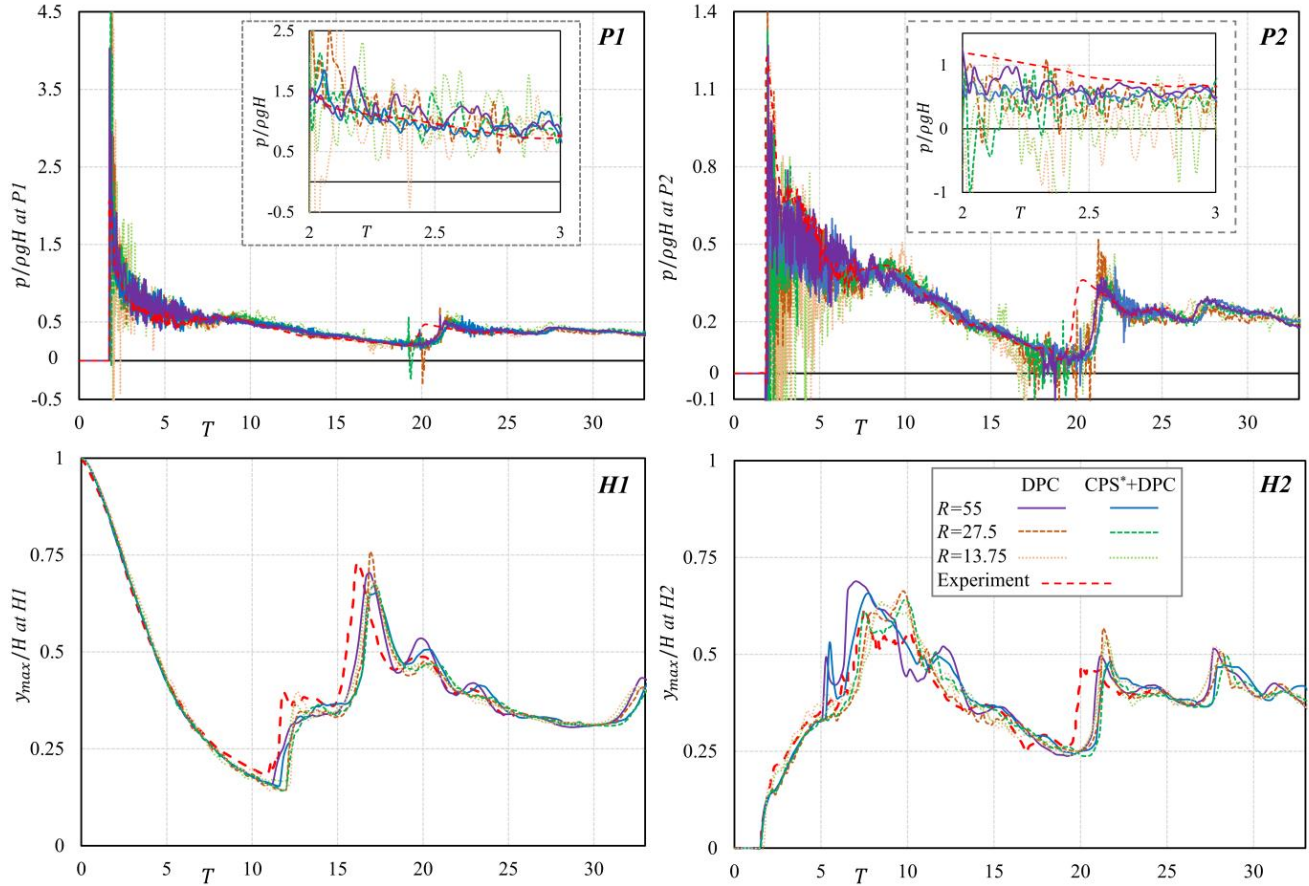

Figure 19- Numerical results (with the DPC and CPS*+DPC techniques and the different spatial resolutions, $R$) versus the experimental measurements from Kleefsman et al. [60]: Shown are the local pressure at points $P1$ and $P2$ (top row) and the depth of water in the y-direction at $H1$ and $H2$ lines, $y_{max}$ (bottom row). The non-dimensional time and the spatial resolutions are given as $T = t\sqrt{g/H}$ and $R = H/l_0$, respectively.

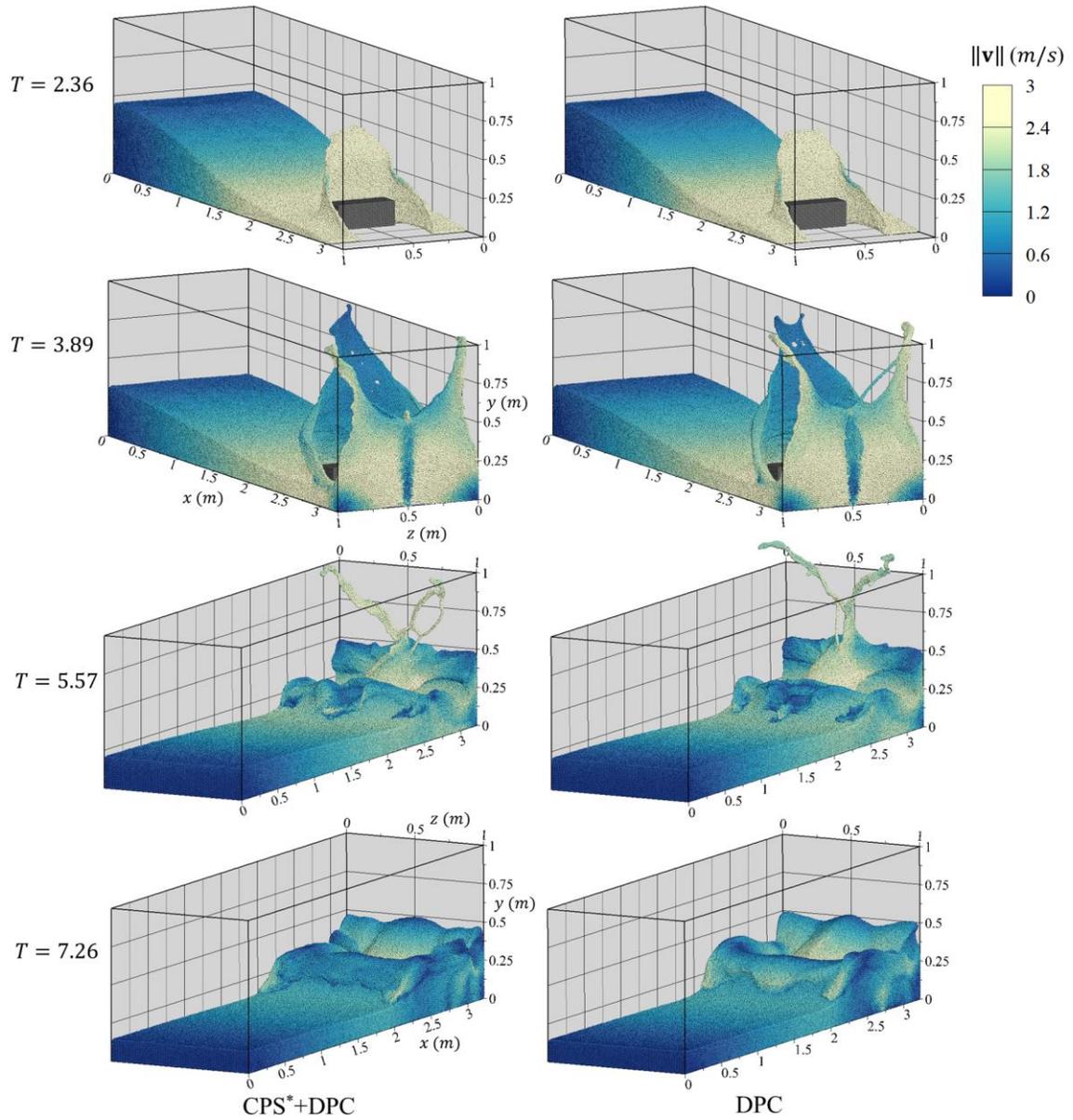

Figure 20- Flow evolution and particles' velocity magnitude for the 3D water dam-break with the rigid obstacle simulated by the model with the CPS$^*$+DPC (left column) and DPC (right column) techniques, where $T = t\sqrt{g/H}$ and $R = 55$.

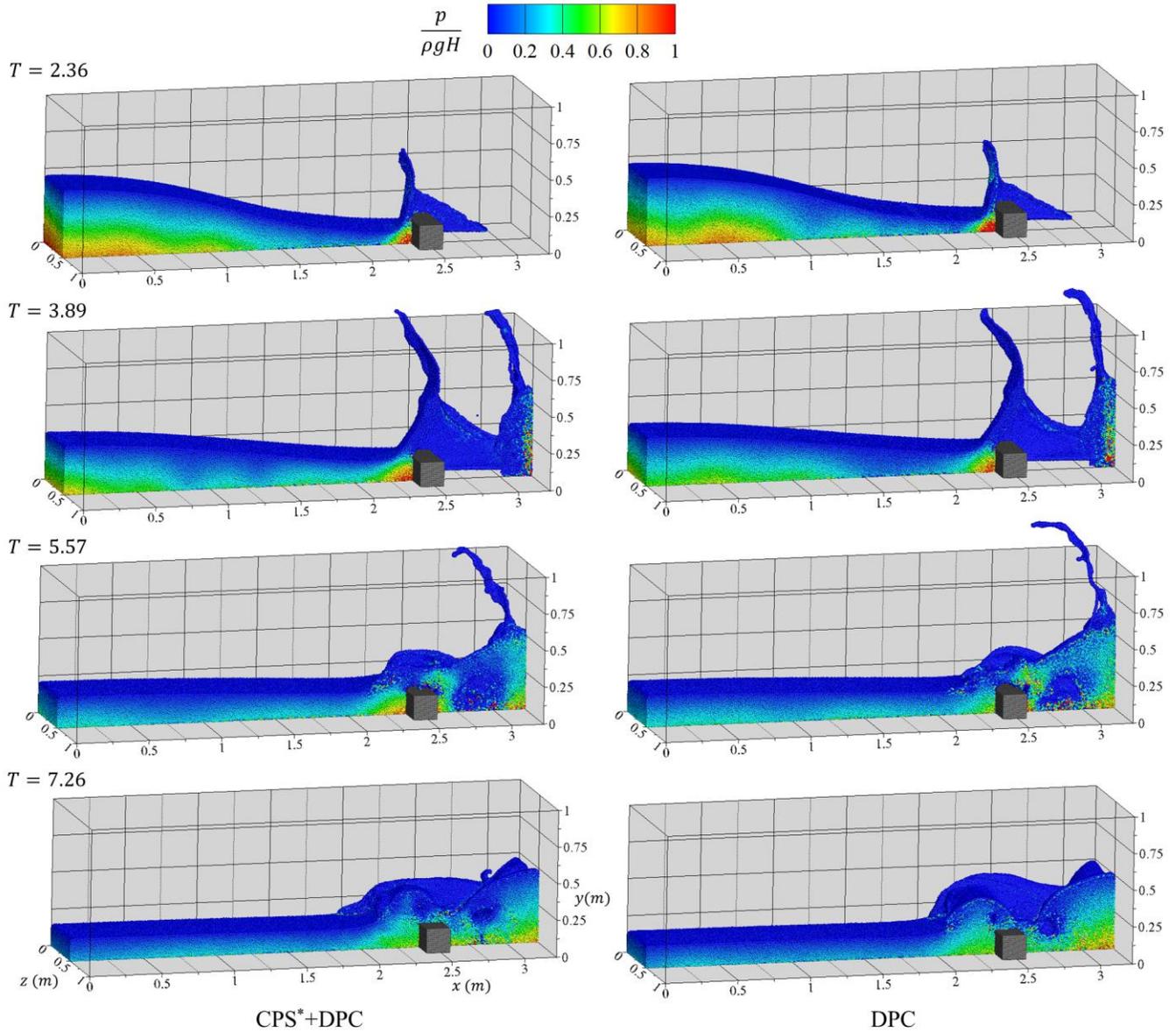

Figure 21- Non-dimensional pressure field of the 3D water dam-break with the rigid obstacle (at Section A-A) during the main impact events simulated by the model with the CPS*+DPC (left column) and DPC (right column) techniques, where $T = t\sqrt{g/H}$ and $R = 55$.

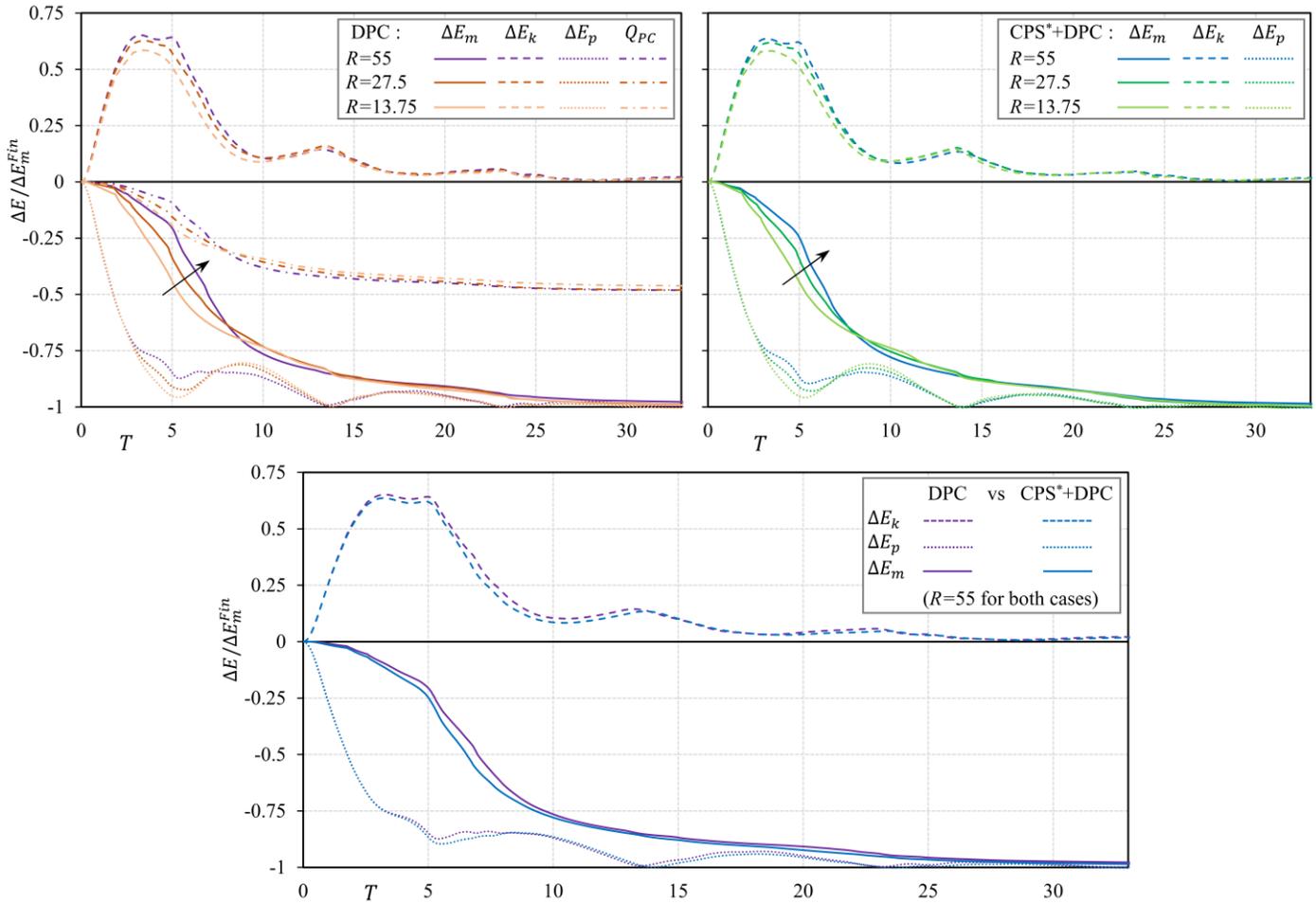

Figure 22- Temporal changes of the energy components of the 3D water dam-break with the obstacle: The top left and right graphs represent the numerical convergence of the 3D model with the DPC and CPS$^*$+DPC techniques (with various spatial resolutions), respectively. The bottom graph compares the energy evolutions with DPC versus CPS$^*$+DPC.

## 5. Conclusion

We developed and validated the three-dimensional EWC-MPS method for simulating violent free-surface flows. In a conservative framework of governing equations, we introduced several enhancement techniques that are shown to be essential for dealing with fluid-fluid and fluid-solid impact events. To include the turbulence shear force, we adopted the SPS scheme with a higher-order gradient operator to estimate the magnitude of the strain rate tensor. Furthermore, we employed a convergent form of the diffusive term in the context of the MPS formulations to reduce high-frequency pressure noises and kernel truncation errors at the boundaries.

We evaluated and compared the impact of the two particle regularization techniques, particle shifting (PS) and pairwise particle collisions (PC), on the accuracy, stability, and convergence of the model. We found that the corrected-PS (in its original form proposed in [21] as CPS+PC) is more effective in reducing the unphysical pressure noises but has a problem with volume conservation in simulating violent free-surface flows (i.e., it expands the volume and deviates the system's potential energy). To resolve the issues with these two regularization techniques, we proposed a dynamic pairwise particle collision and a consistent form of the corrected particle-shifting techniques, denoted as DPC and $CPS^*$+DPC, respectively. The new DPC technique benefits from a dynamic formulation that deals with different states of inter-particle penetration. Unlike the particle-shifting approaches, the DPC is simple to numerically implement, as it does not require any particle classification or free-surface detection. The $CPS^*$+DPC technique consists of a new particle classification algorithm and a modified shifting vector coupled with additional $\delta\hat{\mathbf{v}}$-terms. Moreover, this hybrid approach implements the DPC technique among free-surface and splashed particles in order to avoid unphysical fluid fragmentations.

The overall results confirm the effectiveness and necessity of adopting the proposed regularization techniques for dealing with extreme flow deformations of free-surface violent flows where the potential energy is dominant, and several impact events occur. Both regularization techniques represent accurate flow evolutions of the benchmark cases and eliminate particle-clustering over the fluid domain. Although these techniques require some parameter adjustments, the robustness of both implementations is well demonstrated. The DPC technique, while ensuring numerical stability, results in a smoother pressure field in comparison to the standard PC method. We have shown that the new particle classification algorithm can dynamically detect large domain curvatures and expanded regions required for adjusting the direction and/or magnitude of the particle-shifting vector. The proposed $CPS^*$+DPC technique predicts the final expected potential energy; however, the same model without the $\delta\hat{\mathbf{v}}$-terms still leads to a volume expansion manifested as an increase in the global potential energy of the violent free-surface flow. In all three benchmark cases, the evolution of the energy components showed that the DPC technique is a low-dissipative approach and reduces the calculation costs of the 2D and 3D simulations compared to the $CPS^*$+DPC algorithm that requires the particle classifications and approximating the $\delta\hat{\mathbf{v}}$-terms.

The EWC-MPS method developed and validated in this study can be extended to multiphase flow problems with high-density ratios and complex mechanical behaviors. The robust regularization techniques, along with the modified diffusive term which act independently from the density discontinuity and empirical coefficients, make the method a powerful computational tool for large-scale hydro-environmental and industrial applications. It is also worthwhile to investigate the role of the current developments within the framework of fully incompressible MPS and SPH methods. To do so, further enhancements may include implementing strategies for optimizing the parallel programming (e.g., [58, 61]), acquiring advanced turbulence formulations (e.g., [62]), and consistent solid boundary models (e.g., [54, 63]).

## Acknowledgments

This research was funded by the Natural Sciences and Engineering Research Council of Canada (NSERC) and Polytechnique Montréal, Canada. This study utilized the high-performance computing resources of Compute Canada and Calcul Quebec. The authors also acknowledge Nvidia for supporting this research with their GPU Grant Program.

## Appendix A. The symplectic scheme with the diffusive terms

The second-order symplectic time integration scheme divides the time step into two stages of calculation [56]. In the first stage, the position, velocity, and particle number density of the particles are updated at the midpoint, $t + \Delta t/2$, as follows:

$$
\begin{aligned}
\mathbf{r}_i^{t+\Delta t/2} &= \mathbf{r}_i^t + \frac{\Delta t}{2} \mathbf{v}_i^t \\
\mathbf{v}_i^{t+\Delta t/2} &= \mathbf{v}_i^t + \frac{\Delta t}{2} \left[ -\frac{1}{\rho_i} \langle \nabla p \rangle_i + \mathbf{F}_i + \frac{1}{\rho_i} \langle \nabla . \boldsymbol{\tau} \rangle_i \right]^t \\
n_i^{t+\Delta t/2} &= n_i^t + n_i^t \frac{\Delta t}{2} [-\langle \nabla . \mathbf{v} \rangle_i + D_i^m]^t
\end{aligned}
\qquad (A.1)
$$

With the midpoint density, the equation of state (3) gives the midpoint pressure, $p_i^{t+\Delta t/2}$. In the second stage, with the midpoint values, the right-hand side of the momentum equation estimates the new velocity, and then the new position of the particles is updated at $t + \Delta t$, according to:

$$
\begin{aligned}
\mathbf{v}_i^{t+\Delta t} &= \mathbf{v}_i^t + \Delta t \left[ -\frac{1}{\rho_i} \langle \nabla p \rangle_i + \mathbf{F}_i + \frac{1}{\rho_i} \langle \nabla . \boldsymbol{\tau} \rangle_i \right]^{t+\Delta t/2} \\
\mathbf{r}_i^{t+\Delta t} &= \mathbf{r}_i^{\Delta t/2} + \frac{\Delta t}{2} \mathbf{v}_i^{t+\Delta t}.
\end{aligned}
\qquad (A.2)
$$

Thus, the scheme updates the particle number density (and consequently the pressure, $p_i^{t+\Delta t}$) as follows:

$$n_i^{t+\Delta t} = n_i^{t+\Delta t/2} + n_i^{t+\Delta t/2} \frac{\Delta t}{2}\left[-\langle \nabla . \mathbf{v}\rangle_i^{t+\Delta t} + D_i^{m\,t+\Delta t/2}\right] \quad (A.3)$$

in which the particle number density at the mid-point is used in the $\langle \nabla . \mathbf{v}\rangle_i^{t+\Delta t}$ term.

The solution algorithm is implemented on a GPU accelerated code (Figure A.1). The initialization of the simulation and saving data for the post-processing jobs run on the Computer Processing Unit (CPU) as the host part of the code. Nevertheless, the main temporal loop, including the neighbor search algorithm and the particle regularization techniques, run on the Graphical Processing Unit (GPU) as the device section. The subroutines of this loop are written with the Compute Unified Device Architecture (CUDA) parallel programming language based on C++. This allows the numerical model to use the massive parallelization on the GPU to process a large number of particles simultaneously [58]. Figure A.1 summarizes the numerical implementation of the solution algorithm.

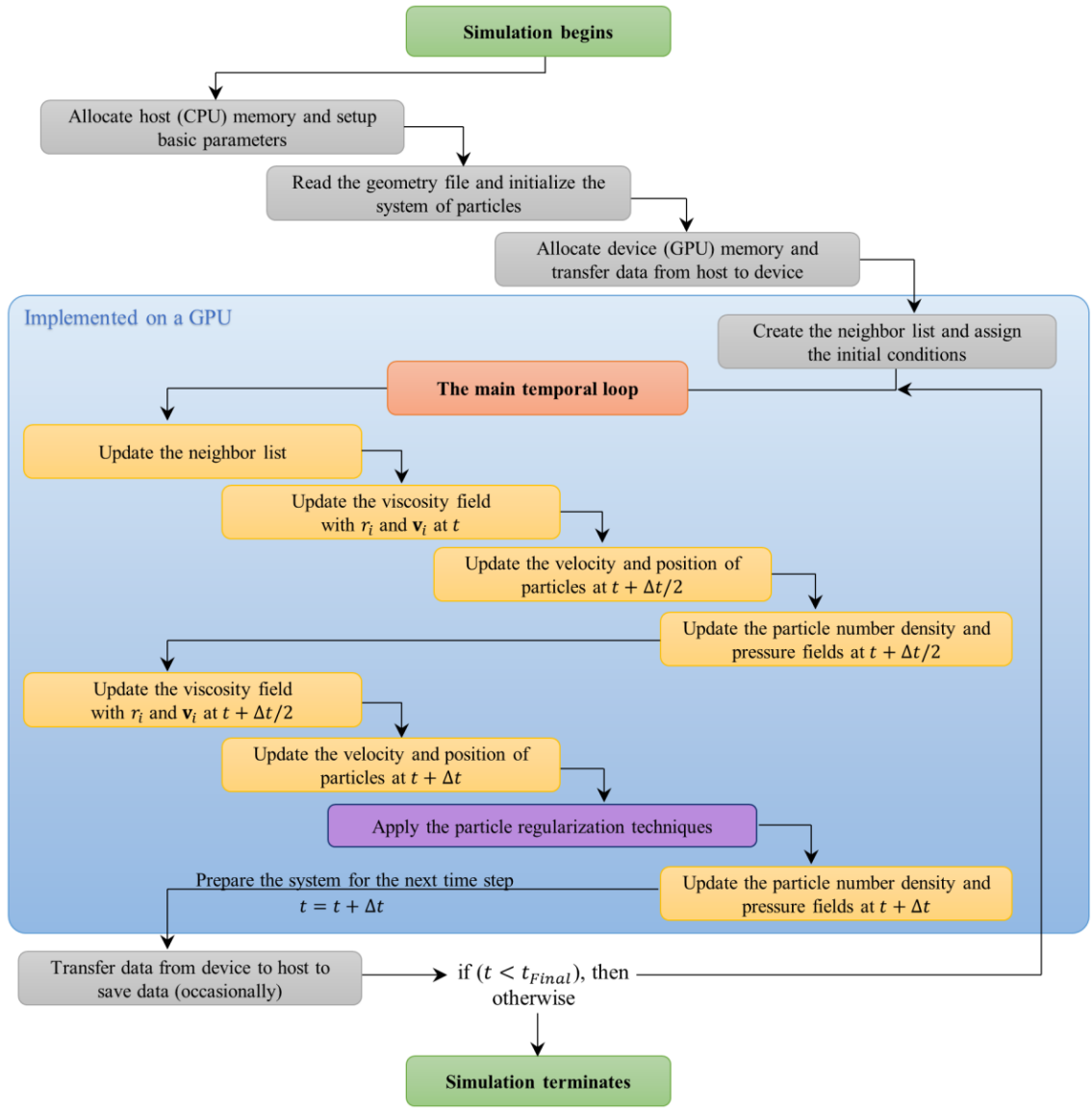

Figure A.1- Flowchart of the solution algorithm based on the symplectic time integration scheme. The main temporal loop of the calculation is implemented on a GPU device with CUDA C++ parallel programming.

## Appendix B. The energy components of the system

By plotting the time evolution of the global energies of the system, we investigate the effect of the particle regularization techniques on the overall mechanical behavior of the system. For this purpose, we calculate the global kinetic energy, $E_k$, the global potential energy, $E_p$, and the global mechanical energy, $E_m$, at each time step, $t$, as follows:

$$E_k^t = \frac{\rho l_0^d}{2} \sum_{i \in \Omega_f} \|\mathbf{v}\|_i^2, \quad E_p^t = -\rho l_0^d \sum_{i \in \Omega_f} \mathbf{g} \cdot \mathbf{r}_i, \quad E_m^t = E_k^t + E_p^t. \quad (B.1)$$

where $\rho l_0^d$ is the constant mass of the fluid particles. The variation of the energies with respect to their initial values, $E^0$, is given as $\Delta E = E^t - E^0$. Furthermore, the total kinetic energy dissipated by the particle collision technique during the simulation time is calculated by:

$$Q_{PC}^t = \sum_{t=0}^{t} (E_k' - E_k)^t \quad (B.2)$$

in which $E_k$ and $E_k'$ are the global kinetic energies before and after updating the velocity of the particles at each time step, respectively.